\documentclass[12pt,a4paper]{iopart}

\usepackage[english]{babel}

\usepackage{iopams}
\usepackage{verbatim}

\usepackage{graphicx}
\usepackage{float}

\usepackage{color}

\usepackage[colorlinks,citecolor=blue,linkcolor=black,urlcolor=blue]{hyperref}

\def\iotabar{\lower3pt\hbox{$\mathchar'26$}\mkern-7mu\iota}
\newcommand {\aplt} {\ {\raise-.5ex\hbox{$\buildrel<\over\sim$}}\ }
\newcommand{\dd}{\mbox{d}}

\newcommand{\eq}[1]{(\ref{#1})}
\newcommand{\bun}{\hat{\mathbf{b}}}
\newcommand{\eun}{\hat{\mathbf{e}}}

\newcommand{\bv}{\mathbf{v}}

\newcommand{\bB}{\mathbf{B}}

\newcommand{\matrixtop}[1]{\buildrel\leftrightarrow\over{#1}}
\newcommand{\matI}{\matrixtop{\mathbf{I}}}

\newcommand{\gsim}{ {\scriptstyle {{_{\displaystyle >}}\atop{\displaystyle \sim}}} }
\newcommand{\dotcross}{ \raise 0.65ex\hbox{${\scriptstyle {{_{\displaystyle \cdot}}\atop\times}}$} }
\newcommand{\crossdot}{ \raise 0.5ex\hbox{${\scriptstyle {{_\times}\atop{\displaystyle \cdot}}}$} }

\newcommand{\cE}{{\cal E}}

\newcommand{\bg}{\mathbf{g}}

\newcommand{\sumsig}{ \raise -1.3ex\hbox{${{\displaystyle \sum}\atop{\scriptstyle \sigma}}$} }
\newcounter{appnumb}

\newcommand{\Ha}{{H_{a}}}
\newcommand{\Hb}{{H_{b}}}

\newcommand{\rcoor}{{r}}

\newcommand{\matM}{\matrixtop{\mathbf{M}}}

\begin{document}

\title[Impact of main ion pressure anisotropy on stellarator impurity transport] {Impact of main ion pressure anisotropy on stellarator impurity transport}


\author{Iv\'an Calvo$^{1}$, F\'elix I. Parra$^{2}$, Jos\'e Luis Velasco$^{1}$ and Jos\' e Manuel Garc\'{\i}a-Rega\~na$^{1}$
}

\vspace{0.2cm}

\address{$^1$Laboratorio Nacional de Fusi\'on, CIEMAT, 28040 Madrid, Spain}
\address{$^2$Rudolf Peierls Centre for Theoretical Physics, University of Oxford, Oxford, OX1 3PU, UK}



\vskip 0.5cm

{\large
\begin{center}
\today
\end{center}
}

\vspace{0.0 cm}

\begin{abstract}
Main ions influence impurity dynamics through a variety of mechanisms; in particular, via impurity-ion collisions. To lowest order in an expansion in the main ion mass over the impurity mass, the impurity-ion collision operator only depends on the component of the main ion distribution that is odd in the parallel velocity. These lowest order terms give the parallel friction of the impurities with the main ions, which is typically assumed to be the main cause of collisional impurity transport. Next-order terms in the mass ratio expansion of the impurity-ion collision operator, proportional to the component of the main ion distribution that is even in the parallel velocity, are usually neglected. However, in stellarators, the even component of the main ion distribution can be very large. In this article, such next-order terms in the mass ratio expansion of the impurity-ion collision operator are retained, and analytical expressions for the neoclassical radial flux of trace impurities are calculated in the Pfirsch-Schl{\"u}ter, plateau and $1/\nu$ regimes. The new terms provide a drive for impurity transport that is physically very different from parallel friction: they are associated to anisotropy in the pressure of the main ions, which translates into impurity pressure anisotropy. It is argued that main ion pressure anisotropy must be taken into account for a correct description of impurity transport in certain realistic stellarator plasmas. Examples are given by numerically evaluating the analytical expressions for the impurity flux.
\end{abstract}


\section{Introduction}
\label{sec:introduction}

Impurity transport is one of the most active research areas in the stellarator community because, as a general rule, impurities are observed to accumulate in the plasma core of these devices~\cite{Burhenn2009}. The accumulation of heavy, highly-charged impurities is extremely harmful for confinement, as it leads to fuel dilution~\cite{Putterich2019} and to intolerable energy losses due to radiation~\cite{Burhenn2009}. Hence, understanding and controlling impurity transport is essential for the success of the stellarator concept as an alternative to tokamaks in the design of future fusion power plants.

In recent years, remarkable progress has been made on the theory of stellarator impurity transport, both analytical~\cite{Braun2010, Helander2017, Calvo2018b, Buller2018} and numerical~\cite{García-Regaña2013, García-Regaña2017, Velasco2018, García-Regaña2018, Mollen2018, Fujita2019}. In \cite{Braun2010}, the radial impurity flux is calculated analytically when all the species in the plasma are collisional. In \cite{Helander2017}, the calculation is given for the case in which impurities are collisional and the main ions have low collisionality. A considerable part of the recent theoretical effort (see \cite{Calvo2018b, Buller2018, García-Regaña2013, García-Regaña2017, Velasco2018, García-Regaña2018, Mollen2018, Fujita2019}) has been related to the extension of neoclassical theory and, consequently, neoclassical codes, to include the effect of the component of the electric field that is tangent to magnetic surfaces in the calculation of radial impurity fluxes. The determination of the radial and tangential components of the electric field is one of the neoclassical mechanisms by which main ion dynamics affects impurity transport. In this article, we will focus on the other obvious mechanism: impurity-ion collisions.

Specifically, we would like to incorporate and quantify a collisional effect usually neglected in the analytical description of neoclassical impurity transport in stellarators: the influence of the pressure anisotropy of the main ions on the dynamics of the impurities via collisions (numerical evidence of the relevance of this effect was given in \cite{Mollen2015} in neoclassical simulations including the exact Landau collision operator). Let us denote by $m_i$ and $m_z$ the mass of the main ions and the impurities, respectively, and by $h_i$ the deviation of the main ion distribution from a Maxwellian distribution. In order to describe collisions between heavy impurities and main ions, the impurity-ion collision operator is expanded in $\sqrt{m_i/m_z}\ll 1$ and typically approximated by the lowest-order term, which only depends on the component of $h_i$ that is odd in the parallel velocity, $h_{i-}$. Next order terms in the mass ratio expansion depend on the component of $h_i$ that is even in the parallel velocity, $h_{i+}$. In tokamaks $h_{i+}$ is small, especially at low main ion collisionality (in this paper, we always assume that the main ions have low collisionality, since this is the situation of interest for fusion purposes), so next-order terms in the $\sqrt{m_i/m_z}\ll 1$ expansion can be safely dropped. However, in stellarators, $h_{i+}$ can be very large in low collisionality plasmas~\cite{Beidler2011, Calvo2017, Calvo2018} due to the large main ion pressure anisotropy, which translates into significant impurity pressure anisotropy that in turn can drive non-negligible radial impurity transport. We present an explicit calculation showing this for impurities in the Pfirsch-Schl\"uter, plateau and $1/\nu$ regimes. In each of these regimes, we will estimate the region of parameter space in which the effect of main ion pressure anisotropy on the impurity flux is relevant.

The rest of the paper is organized as follows. In Section \ref{sec:basic_equations}, we explain the notation and orderings assumed along the paper. We also introduce the trace impurity drift-kinetic equation, including an explicit expression for the impurity-ion collision operator that keeps next-to-lowest-order terms in the $\sqrt{m_i/m_z}\ll 1$ expansion, and hence the effect of $h_{i+}$. In Section \ref{sec:impurities_PS_regime}, we give a detailed calculation of the impurity flux when the impurities are in the Pfirsch-Schl\"uter regime. In Section \ref{sec:impurities_plateau_regime}, the case of impurities in the plateau regime is worked out. In Section \ref{sec:impurities_1overnu_regime}, the neoclassical impurity flux is estimated assuming that the impurities are in the $1/\nu$ regime. In Section \ref{sec:numerical_evaluation}, we evaluate numerically the expressions derived for the impurity flux and illustrate the effect of main ion pressure anisotropy on neoclassical impurity transport in realistic stellarator plasmas. Section \ref{sec:conclusions} contains the conclusions.

\section{Impurity drift-kinetic equation including main ion pressure anisotropy effects}
\label{sec:basic_equations}

In this section we define notation and explain the orderings assumed in the subsequent calculations. We also present the general form of the impurity-drift kinetic equation that will be solved in different regimes in the next sections. In particular, we provide an explicit expression for the impurity-ion collision operator expanded in $\sqrt{m_i/m_z} \ll 1$ to sufficiently high order that it includes the influence of the pressure anisotropy of the main ions.

\subsection{Phase-space coordinates}
\label{sec:phase-space_coordinates}

Before introducing the drift-kinetic equation, we discuss the coordinates that we will employ to locate points on phase space. As velocity coordinates, we take the total energy per mass unit
\begin{equation}
\cE = \frac{v^2}{2} + \frac{Z_z e}{m_z}\varphi,
\end{equation}
the magnetic moment
\begin{equation}
\mu = \frac{v_\perp^2}{2 B},
\end{equation}
the sign of the parallel velocity $\sigma = v_{||} / |v_{||}| = \pm 1$, with
\begin{equation}
v_{||} = \sigma \sqrt{2\left(\cE -\mu B - 
\frac{Z_z e}{m_z} \varphi
\right)} \, ,
\end{equation}
and the gyrophase, $\phi$. Here, $v$ is the magnitude of the velocity $\bv$, $Z_z e$ is the impurity  charge, $e$ is the proton charge, $\varphi$ is the electrostatic potential, $B$ is the magnitude of the magnetic field $\bB$ and $v_\perp^2 = v^2 - v_{||}^2$. The explicit expression of velocity integrals in these coordinates reads
\begin{equation}\label{eq:Jacobian_velocity_space}
\fl
\int (\cdot)\dd^3v \equiv \sum_\sigma\int_{Z_z e \varphi / m_z}^\infty
\dd \cE
\int_0^{B^{-1}(\cE - {Z_z e \varphi / m_z)}} \dd \mu \int_0^{2\pi} \dd \phi \, \frac{B}{|v_{||}|} (\cdot).
\end{equation}

As space coordinates we take $r$, $\theta$ and $\zeta$, where $r$, the local minor radius, is a flux surface label, and $\theta$ and $\zeta$ are poloidal and toroidal angles, respectively. We write the electrostatic potential $\varphi$ as
\begin{equation}
\varphi(r,\theta,\zeta)= \varphi_0(r) + \varphi_1(r,\theta,\zeta),
\end{equation}
where we assume $e \varphi_0 / T \sim 1$ and $|\varphi_1| \ll |\varphi_0|$ (below, we will be more precise about the ordering assumed for the size of $\varphi_1$). The temperatures of the main ions and of the impurities are assumed to be equal and constant on flux surfaces, and denoted by $T(r)$.

In these coordinates, the flux-surface average of any function $f(r,\theta,\zeta)$ is defined by
\begin{equation}\label{eq:def_fluxsurfaceaverage}
\langle f \rangle (\rcoor) = V'(\rcoor)^{-1}\int_{0}^{2\pi} \dd\theta \int_{0}^{2\pi} \dd\zeta
\sqrt{g_{_V}} \, f(r,\theta,\zeta),
\end{equation}
where $\sqrt{g_{_V}} = [(\nabla\rcoor\times\nabla\theta)\cdot\nabla\zeta]^{-1} > 0$ is the volume element and
\begin{equation}\label{eq:def_volume}
V(\rcoor_0) = \int_0^{r_0}\dd\rcoor \int_{0}^{2\pi} \dd\theta \int_{0}^{2\pi} \dd\zeta
\sqrt{g_{_V}}
\end{equation}
is the volume enclosed by the surface labeled by $\rcoor_0$. In this paper, primes stand for derivatives with respect to $r$.

\subsection{Impurity drift-kinetic equation}
\label{sec:impurity_DKE_subsection}

Throughout this paper we use the trace impurity approximation,
\begin{equation}\label{eq:trace_impurity_approximation}
\frac{Z_z^2 n_z}{Z_i^2 n_i} \sqrt{\frac{m_z}{m_i}}\ll 1,
\end{equation}
where $Z_i$ is the main ion charge number and $n_i(r)$ is the lowest-order density of the main ions, which is a flux function, whereas the lowest order impurity density, $n_z$, is not a flux function if $Z_z e \varphi_1 / T$ is large enough (see below). When \eq{eq:trace_impurity_approximation} is satisfied, impurity-impurity collisions can be neglected with respect to impurity-ion collisions (impurity-electron collisions can be neglected against impurity-ion collisions due to the small mass of the electrons relative to the mass of the main ions).

We assume that the characteristic impurity gyroradius is much smaller than the characteristic size of the stellarator, $\rho_{z*} := v_{tz}/\Omega_z R_0 \ll 1$, that the $\mathbf{E} \times \mathbf{B}$ drift, with $\mathbf{E} = - \nabla\varphi$, is small compared to the impurity thermal speed (see below a more detailed discussion on this approximation) and that the typical collision frequency between impurities and ions and the characteristic parallel streaming time are comparable, $\nu_{zi*} := R_0 \nu_{zi}/v_{tz} \sim 1$ (we will perform a subsidiary expansion in small and large $\nu_{zi*}$ in what follows). Here $v_{tz} = \sqrt{2T_z/m_z}$ is the impurity thermal speed, $\Omega_z = Z_z e B/m_z$ is the impurity gyrofrequency, $R_0$ is the stellarator major radius,
\begin{equation}
\nu_{zi} = \frac{Z_z^2 Z_{i}^2 e^4 n_i m_i^{1/2} \ln\Lambda}{6\sqrt{2}\pi^{3/2}\epsilon_0^2 m_z T^{3/2}}
\,
\end{equation}
is the impurity-ion collision frequency, $\epsilon_0$ is the vacuum permittivity and $\ln\Lambda$ is the Coulomb logarithm.

Using these assumptions, the impurity distribution $F_z (r, \theta, \zeta, \cE, \mu, \sigma)$ can be shown to be independent of the gyrophase $\phi$ to the relevant order~\cite{Calvo13}, and the particle motion can be split into fast thermal motion along magnetic field lines and slow drifts across, giving the trace impurity drift-kinetic equation
\begin{equation}\label{eq:DKE_impurities_full_f}
v_{||}\bun\cdot\nabla F_z + \bv_d \cdot\nabla F_z = C_{zi}[F_z,F_i],
\end{equation}
where $C_{zi}[F_z,F_i]$ is the impurity-ion collision term, $F_i$ is the main ion distribution and $\bv_d = \bv_M + \bv_E$ is the sum of the magnetic drift
\begin{equation}
\bv_M = \frac{v_{||}^2}{\Omega_z}\bun\times(\bun\cdot\nabla\bun) + \frac{\mu }{\Omega_z}\bun\times\nabla B
\end{equation}
and the $\mathbf{E} \times \mathbf{B}$ drift
\begin{equation}
\bv_E = \frac{1}{B} \bun\times \nabla(\varphi_0 + \varphi_1).
\end{equation}

Of all our assumptions, the small $\mathbf{E}\times \mathbf{B}$ drift approximation is the one that is more likely to fail for highly charged impurities, because for a typical electric field $|\mathbf{E}| \sim |\nabla \varphi_0| \sim T/(ea)$, with $a$ the stellarator minor radius, the ratio of the term $B^{-1}(\bun\times\nabla\varphi_0)\cdot\nabla F_{z}$ to the parallel streaming term $v_{||}\bun\cdot\nabla F_z$ has a size
\begin{equation} \label{eq:Machnumber}
\frac{|\mathbf{E}|}{B v_{tz}}\frac{1}{\epsilon} \sim \sqrt{\frac{m_z}{m_i}}\, \frac{\rho_{i*}}{\epsilon^2},
\end{equation}
where $\rho_{i*} = v_{ti}/(\Omega_i R_0)$ is the normalized gyroradius of the main ions, $v_{ti} = \sqrt{2T / m_i}$ is the main ion thermal speed, $\Omega_i = Z_i e B / m_i$ is the main ion gyrofrequency and $\epsilon = a/R_0$ is the stellarator inverse aspect ratio. If $|\mathbf{E}|/(\epsilon B v_{tz})$ is of order unity, the $\mathbf{E}\times \mathbf{B}$ drift term becomes comparable to the parallel streaming term, and the perpendicular drifts are modified by Coriolis and centrifugal forces~\cite{Bernstein1985}. Thus, our expression will only be valid in the cases when the estimate~(\ref{eq:Machnumber}) is small, which, for sufficiently heavy impurities, amounts to requiring small radial electric field. We do not consider this a major constraint for the main result of this article because, as we will see, the main ion pressure anisotropy contribution becomes important for small radial electric fields.

Let us expand $F_z$ in the small parameter $\rho_{z*}$,
\begin{equation}
F_z = F_{z0} + F_{z1} + \dots,
\end{equation}
with $F_{z1} \sim \rho_{z*} F_{z0}$. The lowest order terms in \eq{eq:DKE_impurities_full_f} give
\begin{equation}\label{eq:DKE_impurities_full_f_lowest_order}
v_{||}\bun\cdot\nabla F_{z0} = C_{zi}[F_{z0},F_{Mi}],
\end{equation}
where we are assuming that, to lowest order in the expansion, the main ion distribution is a Maxwellian distribution
\begin{equation}
F_{Mi}  = 
n_i(\rcoor)\left(\frac{m_i}{2\pi T(\rcoor)}\right)^{3/2}
\exp\left(-\frac{m_i v^2}{2T(\rcoor)}\right).
\end{equation}

 \ref{sec:C_zi} contains a detailed calculation of the lowest order terms of the linearized impurity-ion collision operator expanded in $\sqrt{m_i/m_z}$. From the results in \ref{sec:C_zi}, one can infer (as a particular case) that the collisions of heavy impurities with a Maxwellian background of main ions are given by
\begin{equation}\label{eq:coll_impurities_Maxwellian_ions}
C_{zi}[F_{z0},F_{Mi}] =
\nu_{zi}
\frac{T}{m_z}\nabla_v\cdot\left(
\nabla_v F_{z0} + \frac{m_z \bv}{T} F_{z0}
\right).
\end{equation}

Multiplying \eq{eq:DKE_impurities_full_f_lowest_order} (with the collision operator given in \eq{eq:coll_impurities_Maxwellian_ions}) by $\exp(m_z {\cal E}/T(r)) F_{z0}$, integrating over velocities and flux-surface averaging, we get
\begin{equation}\label{eq:DKE_impurities_full_f_lowest_order_aux}
\fl
- \left
\langle
\exp\left(\frac{Z_ze(\varphi_0 + \varphi_1)}{T}\right)
\int \exp\left(-\frac{m_z v^2}{2T}\right)\left| \nabla_v \left(
\exp\left(\frac{m_z v^2}{2T}\right) F_{z0}
\right)\right|^2
\dd^3 v
\right
\rangle = 0,
\end{equation}
which implies
\begin{equation}
\nabla_v \left(
\exp\left(\frac{m_z v^2}{2T}\right) F_{z0}
\right) = 0.
\end{equation}
That is, $F_{z0} = F_{Mz}$, where $F_{Mz}$ is a Maxwellian distribution with the same temperature as $F_{Mi}$. Going back to \eq{eq:DKE_impurities_full_f_lowest_order} and using that $C_{zi}[F_{Mz},F_{Mi}] = 0$, we obtain
\begin{equation}
v_{||}\bun\cdot\nabla F_{Mz} = 0,
\end{equation}
which, on an ergodic surface, implies
\begin{equation}
\fl
F_{Mz}(r,\cE)=
{\eta}(r)
\left(
 \frac{m_z}{2\pi T(r)}
 \right)^{3/2}
 \exp\left(\frac{Z_z e\varphi_0(r)}{T(r)}\right)
 \exp\left(
-\frac{m_z \cE}{T(r)}
\right).
\end{equation}
The flux function $\eta(r)$ is related to the density of the Maxwellian distribution,
\begin{equation}
n_z := \int F_{Mz}\dd^3 v,
\end{equation}
by the relation
\begin{equation}\label{eq:def_n_z}
n_z(r,\theta,\zeta) = {\eta}(r) \exp\left(-\frac{Z_z e \varphi_1(r,\theta,\zeta)}{T(r)}\right).
\end{equation}

The next-order terms of the drift-kinetic equation \eq{eq:DKE_impurities_full_f} in the $\rho_{z*}$ expansion give an equation for $F_{z1}$,
\begin{equation}\label{eq:DKE_impurities}
v_{||}\bun\cdot\nabla F_{z1} + \bv_d \cdot\nabla r \Upsilon F_{Mz} = C^{(\ell)}_{zi}[F_{z1};h_i].
\end{equation}
Here,
\begin{equation}\label{def:Upsilon}
\Upsilon = \frac{\eta'}{\eta} + \frac{Z_ze\varphi'_0}{T}
+\frac{T'}{T}\left(
\frac{m_z\cE}{T} - \frac{Z_z e \varphi_0}{T} - \frac{3}{2}
\right)
\end{equation}
is a combination of the thermodynamic forces, $h_i$ is the non-adiabatic component of the deviation of the main ion distribution from a Maxwellian distribution,
\begin{equation}\label{eq:radial_drift}
\bv_d \cdot \nabla\rcoor
=
\frac{v_{||}}{\Omega_z}\nabla\cdot\left(v_{||}\bun\times\nabla r\right)
\end{equation}
is the radial drift, $\bv_d \cdot \nabla\rcoor = \bv_M \cdot \nabla\rcoor + \bv_E \cdot \nabla\rcoor$, and $C^{(\ell)}_{zi}[F_{z1};h_i]$ is the linearization of $C_{zi}$ around $F_{Mz}$ and $F_{Mi}$,
\begin{eqnarray}\label{eq:C_zi_lin_expanded_in_mass_ratio}
\fl
C^{(\ell)}_{zi}[F_{z1};h_{i}] =
\nu_{zi}
{\cal K}F_{z1}
+
\nu_{zi}\frac{m_z}{T}
\left(
A v_{||}  - \tr(\matM)  + \frac{m_z}{T} \matM \, : \bv\bv
\right)
F_{Mz},
\end{eqnarray}
where the operator ${\cal K}$ is defined by
\begin{equation}
{\cal K}F_{z1} := \frac{T}{m_z}
\nabla_v\cdot
\left(
F_{Mz}
\nabla_v
\left(\frac{F_{z1}}{F_{Mz}}\right)
\right).
\end{equation}

The quantities $A$ and $\matM$, that depend on the main ions, are given by
\begin{eqnarray}\label{eq:def_vectorA}
A
=
\frac{3\sqrt{\pi\,}T^{3/2}}{\sqrt{2}n_i m_i^{3/2}}
\int
\frac{v_{||}}{v^3}
h_{i}(\bv)
\dd^3v,
\end{eqnarray}
\begin{eqnarray}\label{eq:matrix_M}
\matM
\, =
- \frac{3\sqrt{\pi}\, T^{5/2}}{2\sqrt{2}\, n_i m_z m_i^{3/2}}
\int
\nabla_{v'}\nabla_{v'}\nabla_{v'}v' 
\cdot
\nabla_{v'}\left(\frac{h_{i}(\bv')}{F_{Mi}}\right) F_{Mi}
\dd^3v'.
\end{eqnarray}
A detailed derivation of expression \eq{eq:C_zi_lin_expanded_in_mass_ratio} can be found in \ref{sec:C_zi}. It will often be useful to employ that isotropy of the collision operator in velocity space implies
\begin{equation}\label{eq:Mvv}
{\matM} : \bv\bv = {\matM} : \frac{1}{2\pi}\int \bv\bv \dd\phi = {\matM} :
\left[
v_{||}^2\bun\bun + \frac{v_\perp^2}{2}\left(\matI - \bun\bun\right)
\right]
\end{equation}

From \eq{eq:def_vectorA} and \eq{eq:matrix_M} it is clear that only $h_{i-}$ enters $A$ and only $h_{i+}$ enters $\matM$. Since the trace impurity approximation is made, the calculation of $h_{i}$ and $\varphi_1$ does not involve the impurities, and therefore $h_i$ and $\varphi_1$ enter as known functions in \eq{eq:DKE_impurities}. Let us explicitly write equation \eq{eq:DKE_impurities} with the collision operator \eq{eq:C_zi_lin_expanded_in_mass_ratio},
\begin{eqnarray}\label{eq:DKE_impurities_more_explicit}
v_{||}\bun\cdot\nabla F_{z1} - \nu_{zi}
{\cal K}F_{z1}
 =
 \nonumber\\[5pt]
 \hspace{1cm}
 - \bv_d \cdot\nabla r \Upsilon F_{Mz} +
\nu_{zi}\frac{m_z}{T}
\left(
A v_{||}  - \tr(\matM)  + \frac{m_z}{T} \matM \, : \bv\bv
\right)
F_{Mz}.
\end{eqnarray}

In terms of the solution of the impurity drift-kinetic equation \eq{eq:DKE_impurities_more_explicit}, the radial impurity flux across the surface $r$ reads
\begin{equation}\label{eq:def_Gamma_z}
\Gamma_z[F_{z1}] = \left\langle \int \bv_d \cdot \nabla \rcoor F_{z1} \dd^3 v \right\rangle.
\end{equation}
In this paper we will calculate \eq{eq:def_Gamma_z} for several impurity collisionality regimes. Since equation \eq{eq:DKE_impurities_more_explicit} is linear, we can treat the effect of its source terms independently. We will denote by $\Gamma_{z,a}$ the radial flux produced by the source terms containing the radial drift and $A$ (i.e., $\Gamma_{z,a}$ is the impurity flux obtained by neglecting main ion pressure anisotropy), and by $\Gamma_{z,b}$ the radial flux produced by the source terms containing $\matM$ (i.e., $\Gamma_{z,b}$ is the impurity flux due to main ion pressure anisotropy). By comparing the sizes of $\Gamma_{z,a}$ and $\Gamma_{z,b}$ we will learn when main ion pressure anisotropy is non-negligible.

So far, we have not made explicit assumptions about the aspect ratio of the stellarator. In the next subsection, we expand in large aspect ratio and derive the size and scaling with aspect ratio of $A$ and $\matM$.

\subsection{Estimates for the size of $A$ and $\matM$ in large aspect ratio stellarators}
\label{sec:large_aspect_ratio}

When $\epsilon = a/R_0 \ll 1$, one can write
\begin{equation}\label{eq:ordering_B}
B(r,\theta,\zeta) = B_0 + B_1(r,\theta,\zeta),
\end{equation}
where $B_0$ is constant on the flux surface and $B_1 = O(\epsilon B_0)$. As for $\Upsilon$, we assume
\begin{equation}\label{eq:def_Upsilon}
\Upsilon \sim \frac{Z_z}{a}.
\end{equation}
The factor $Z_z$ in the previous expression is easy to understand looking at \eq{def:Upsilon} and noting that $e\varphi'_0/T \sim 1/a$. For trace impurities, equilibrium conditions imply $\eta' / \eta \sim Z_z/a$ as well.

We also need to order $\varphi_1$ with respect to $\epsilon$. We take
\begin{equation}\label{eq:ordering_varphi1}
\frac{Z_z e \varphi_1}{T} \sim \epsilon,
\end{equation}
which is the size of $\varphi_1$ that makes the radial components of the magnetic and $\mathbf{E} \times \mathbf{B}$ drifts comparable. In \cite{Calvo2018b}, it was analytically proven that when $Z_z e |\varphi_1| / T \gsim \epsilon$, the effect of $\varphi_1$ on the radial neoclassical flux of collisional impurities matters.

It will be important to know the size of $A$ and $\matM$ when $\epsilon \ll 1$. An analytical expression for $A$ was given in \cite{Calvo2018b} for low collisionality main ions, which implies
\begin{equation}
A \sim \frac{1}{\epsilon^2} \rho_{i*} v_{ti}.
\end{equation}
Importantly, from the expression in \cite{Calvo2018b}, one can see that
\begin{equation}\label{eq:A_for_small_epsilon}
A = A_0 + A_1,
\end{equation}
where $A_0$ is constant on the flux surface and $A_1(r,\theta,\zeta) = O(\epsilon A_0) = O(\epsilon^{-1} \rho_{i*} v_{ti})$. An explicit expression for $A_0$ will be given in subsection \ref{sec:solution_dke_Gz2}.

We proceed to estimate the size of $\matM$ for low collisionality main ions. It is useful to employ $v$, $\lambda = v_\perp^2/(Bv^2)$, $\sigma$ and $\phi$ as velocity coordinates for the main ions. In these coordinates, the parallel velocity is written as
\begin{equation}
v_{||} = \sigma v \sqrt{1-\lambda B} \, .
\end{equation}

Since $h_i$ and $F_{Mi}$ are independent of the gyrophase, we have
\begin{eqnarray}
\fl
\nabla_{v}\nabla_{v}\nabla_{v} v\cdot\nabla_{v}
\left(
\frac{h_{i}(\bv)}{F_{Mi}(\bv)}
\right)
=
\nabla_{v}\nabla_{v}\nabla_{v} v\cdot\nabla_{v} v \, \partial_v
\left(
\frac{h_{i}(\bv)}{F_{Mi}(\bv)}
\right)
\nonumber\\[5pt]
\fl\hspace{1cm}
+
\nabla_{v}\nabla_{v}\nabla_{v} v\cdot\nabla_{v} \lambda \, \partial_\lambda
\left(
\frac{h_{i}(\bv)}{F_{Mi}(\bv)}
\right).
\end{eqnarray}

Employing
\begin{equation}
(\nabla_{v}\nabla_{v}\nabla_{v} v)_{ijk}
=
\frac{3}{v^5}v_i v_j v_k -\frac{1}{v^3}
\left(
\delta_{ij}v_k + \delta_{jk}v_i + \delta_{ki}v_j
\right)
\end{equation}
to derive the identity
\begin{equation}
\nabla_{v}\nabla_{v}\nabla_{v} v\cdot\mathbf{Y}
=
\frac{3}{v^5}\bv\bv\bv\cdot\mathbf{Y}
-
\frac{1}{v^3}
\left(\bv\cdot\mathbf{Y}\matI + \bv\mathbf{Y} + \mathbf{Y}\bv\right),
\end{equation}
valid for any vector $\mathbf{Y}$, we find
\begin{equation}\label{eq:nabla3vnablav}
\nabla_{v}\nabla_{v}\nabla_{v} v\cdot\nabla_{v} v
=
-\frac{1}{v^2}
\left(
\matI - \frac{\bv\bv}{v^2}
\right)
\end{equation}
and
\begin{equation}\label{eq:nabla3vnablalambda}
\nabla_{v}\nabla_{v}\nabla_{v} v\cdot\nabla_{v} \lambda
=
-\frac{1}{v^3}
\left(
\bv\nabla_v\lambda + \nabla_v\lambda\bv
\right),
\end{equation}
where we have used
\begin{equation}
\nabla_v v = \frac{1}{v}\bv
\end{equation}
and 
\begin{equation}
\nabla_v\lambda = \frac{2}{B}
\frac{v_{||}}{v^2}\left(
\frac{v_{||}}{v^2}\bv - \bun
\right).
\end{equation}

The gyroaverage of \eq{eq:nabla3vnablav} and \eq{eq:nabla3vnablalambda} gives
\begin{eqnarray}
\fl
\frac{1}{2\pi}\int_0^{2\pi}\nabla_{v}\nabla_{v}\nabla_{v} v\cdot\nabla_{v} v \,
\dd\phi
=
\frac{1}{v^2}
\left(
\frac{v_\perp^2}{2v^2} - 1
\right)
\left(\matI - \bun\bun\right)
+
\frac{1}{v^2}\left(
\frac{v_{||}^2}{v^2} - 1
\right)\bun\bun
\end{eqnarray}
and
\begin{eqnarray}
\fl
\frac{1}{2\pi}\int_0^{2\pi}\nabla_{v}\nabla_{v}\nabla_{v} v\cdot\nabla_{v}\lambda \,
\dd\phi
=
\frac{2v_{||}^2 v_\perp^2}{Bv^7}
\left(
3\bun\bun - \matI
\right)
.
\end{eqnarray}
In this paper, we always assume that the main ions have low collisionality. We can take $h_{i+}$, the component of $h_i$ that is even in $v_{||}$, to be zero for passing particles~\cite{Calvo2017, Calvo2018, Helander2017b}. Trapped trajectories are defined by $\lambda\in[B_{\rm max}^{-1}, B_{\rm min}^{-1}]$, where $B_{\rm max}$ and $B_{\rm min}$ are, respectively, the maximum and minimum values of $B$ on the flux surface. Due to \eq{eq:ordering_B}, the region of phase space corresponding to trapped particles has a size $O(\epsilon B_0^{-1})$ in the coordinate $\lambda$. Hence, for trapped particles, $v_{||}\sim \epsilon^{1/2}v_{ti}$, $v\sim v_\perp\sim v_{ti}$, and we can approximate
\begin{eqnarray}
\frac{1}{2\pi}\int_0^{2\pi}\nabla_{v}\nabla_{v}\nabla_{v} v\cdot\nabla_{v} v \,
\dd\phi
\simeq
-\frac{1}{2v^2}\left(\matI + \bun\bun\right)
\end{eqnarray}
and
\begin{eqnarray}
\frac{1}{2\pi}\int_0^{2\pi}\nabla_{v}\nabla_{v}\nabla_{v} v\cdot\nabla_{v}\lambda \,
\dd\phi
\simeq
\frac{2v_{||}^2}{Bv^5}
\left(
3\bun\bun - \matI
\right)
.
\end{eqnarray}

Finally, for $\epsilon\ll 1$, the matrix $\matM$ has the form
\begin{eqnarray}\label{eq:M_small_epsilon}
\matM
\, =
- \frac{3\sqrt{\pi}\, T^{5/2}}{2\sqrt{2}\, n_i m_z m_i^{3/2}}
\int_0^\infty
\dd v
\int_{B_{\rm max}^{-1}}^{B^{-1}}
\dd\lambda
\frac{\pi B}{|v_{||}|}
\Bigg[
2\left(3\bun\bun - \matI \right) h_{i+}
\nonumber\\[5pt]
\hspace{1cm}
-
v
\left(\matI + \bun\bun \right)
\partial_v\left(\frac{h_{i+}}{F_{Mi}}\right)F_{Mi}
\Bigg],
\end{eqnarray}
where we have used that in coordinates $\{v,\lambda,\phi\}$
\begin{equation}
\int (\cdot) \, \dd^3v
\equiv
\sum_\sigma
\int_0^\infty\dd v
\int_0^{B^{-1}}\dd\lambda
\int_0^{2\pi}\dd\phi \, \frac{v^3 B}{2|v_{||}|} \, (\cdot)
\end{equation}
and an integration by parts in $\lambda$ has been performed in order to obtain the form of the first term on the right side of \eq{eq:M_small_epsilon}. Expression \eq{eq:M_small_epsilon} implies
\begin{equation}\label{eq:sizeM}
\tr(\matM)  \sim \frac{m_z}{T} \matM \, : \bv\bv
\sim \epsilon^{1/2} \frac{T}{m_z}\frac{h_{i+}}{F_{Mi}} \, .
\end{equation}
If the main ions are in the $1/\nu$ regime, defined by $\nu_{ii*}\ll \epsilon^{3/2}$, then~\cite{Calvo2018}
\begin{equation}\label{eq:hi_over_FMi_1overnu}
\frac{h_{i+}}{F_{Mi}} \sim \frac{\rho_{i*}}{\nu_{ii*}},
\end{equation}
where $\nu_{ii*} = R_0\nu_{ii}/v_{ti}$ is the main ion collisionality and $\nu_{ii}$ is the ion-ion collision frequency. Then, the size of the matrix $\matM$ is
\begin{equation}\label{eq:sizeM_1overnu}
\tr(\matM)  \sim \frac{m_z}{T} \matM \, : \bv\bv
\sim \epsilon^{1/2} \frac{T}{m_z}\frac{\rho_{i*}}{\nu_{ii*}}.
\end{equation}

\section{Impurities in the Pfirsch-Schl\"uter regime}
\label{sec:impurities_PS_regime}

The impurities are said to be in the Pfirsch-Schl\"uter regime when they are collisional; i.e.
\begin{equation}
\nu_{zi*} := R_0 \nu_{zi} / v_{tz} \gg 1.
\end{equation}
In order to solve \eq{eq:DKE_impurities_more_explicit}, we write $F_{z1} = H_a + H_b$, where $H_a$ and $H_b$ are the solutions of
\begin{eqnarray}\label{eq:dke_z_old}
\fl
v_{||}\bun\cdot\nabla H_a 
-\nu_{zi}
{\cal K} H_a =
- \bv_d\cdot\nabla \rcoor \partial_r F_{Mz}
+
\nu_{zi}\frac{m_z A v_{||}}{T} F_{Mz} 
\end{eqnarray}
and
\begin{eqnarray}\label{eq:dke_z_new}
\fl
v_{||}\bun\cdot\nabla H_b
-
\nu_{zi}
{\cal K} H_b=
\nu_{zi} \frac{m_z}{T}
\left(
\frac{m_z}{T}
\matM \, : \bv\bv - \tr(\matM)
\right) 
F_{Mz}.
\end{eqnarray}
An explicit expression for the radial flux due to $H_a$ can be found in equation (25) of reference  \cite{Calvo2018b}. Assuming the ordering \eq{eq:ordering_varphi1} for $\varphi_1$, the result of \cite{Calvo2018b} gives
\begin{equation}\label{eq:Gammaa_PS}
\Gamma_{z,a} := \Gamma_z[H_a] \sim \frac{m_z n_z T \nu_{zi}}{Z_z e^2 B^2 a}.
\end{equation}

In this section, we focus on the calculation of $\Gamma_{z,b} := \Gamma_z[H_b]$. For that, we must solve \eq{eq:dke_z_new} in the asymptotic limit $\nu_{zi*} \gg 1$. Subsections \ref{sec:solution_eq_H_2_organization},  \ref{sec:solution_eq_H_2_order_minus1}, \ref{sec:solution_eq_H_2_order_zero} and \ref{sec:solution_eq_H_2_solvability_condition} are devoted to this. In subsection \ref{sec:explicit_expression_Gammab_PS}, an explicit expression for $\Gamma_{z,b}$ is provided. In subsection \ref{sec:comparison_Gammaa_vs_Gammab_PS}, the size of the ratio $\Gamma_{z,b}/\Gamma_{z,a}$ is estimated.

\subsection{Organization of the calculation}
\label{sec:solution_eq_H_2_organization}

We expand $H_b$  in powers of $1/\nu_{zi*} \ll 1$,
\begin{equation}
H_b = H_b^{(0)} + H_b^{(1)} + H_b^{(2)} + \dots,
\end{equation}
with $H_b^{(k)} \propto (1/\nu_{zi*})^{k}$. The terms in \eq{eq:dke_z_new} that scale with $\nu_{zi*}$ yield
\begin{eqnarray}\label{eq:dke_z_new_order_minus1}
\nu_{zi}
{\cal K} H_b^{(0)}
=
\nu_{zi} \frac{m_z}{T}
\left(
\tr(\matM) - \frac{m_z}{T} \matM \, : \bv\bv
\right)F_{Mz}.
\end{eqnarray}
Since
\begin{equation}
\int {\cal K} f \dd^3v = 0
\end{equation}
for any phase-space function $f$, in principle equation \eq{eq:dke_z_new_order_minus1} might have a solvability condition. However, it does not, because the integral over velocities on the right-hand side of \eq{eq:dke_z_new_order_minus1} vanishes.

To next order in $1/\nu_{zi*}$, equation \eq{eq:dke_z_new} gives
\begin{eqnarray}\label{eq:dke_z_new_order_zero}
\nu_{zi}
{\cal K} H_b^{(1)}=
v_{||}\bun\cdot\nabla H_b^{(0)}
.
\end{eqnarray}
This equation does not have any solvability condition either because $H_b^{(0)}$ is even in $v_{||}$ and, therefore, the velocity integral of its right-hand side vanishes.

Finally, from terms in \eq{eq:dke_z_new} that scale as $1/\nu_{zi*}$, we find
\begin{eqnarray}\label{eq:dke_z_new_order_1}
\nu_{zi}
{\cal K} H_b^{(2)}=
v_{||}\bun\cdot\nabla H_b^{(1)}.
\end{eqnarray}
We will not need to solve \eq{eq:dke_z_new_order_1}, but only deal with its solvability condition, which in this case is non-trivial. It is obtained by integrating \eq{eq:dke_z_new_order_1} over velocities, giving
\begin{equation}\label{eq:H2_solvability_condition}
\int \frac{v_{||}}{B} H_b^{(1)}\dd^3 v = Q(r),
\end{equation}
where $Q(r)$ is a flux function to be computed. It is necessary to carry out the calculation to this order in the $1/\nu_{zi*} \ll 1$ expansion to completely determine $H_b^{(0)}$, which gives the dominant contribution of $H_b$ to $\Gamma_{z,b}$ in this collisionality regime; that is,
\begin{equation}\label{eq:Gammab_PS}
\Gamma_{z,b} \simeq \Gamma_z[H_b^{(0)}] .
\end{equation}

In subsection \ref{sec:solution_eq_H_2_order_minus1}, we solve \eq{eq:dke_z_new_order_minus1}. In subsection \ref{sec:solution_eq_H_2_order_zero}, the solution of \eq{eq:dke_z_new_order_zero} is given. In subsection \ref{sec:solution_eq_H_2_solvability_condition}, we work out \eq{eq:H2_solvability_condition}.

\subsection{Solution of \eq{eq:dke_z_new_order_minus1}}
\label{sec:solution_eq_H_2_order_minus1}

\subsubsection{Solution of the homogeneous equation associated to equation \eq{eq:dke_z_new_order_minus1}.}
\label{sec:solution_eq_H_2_order_minus1_homogeneous}

Let us call $H_{b,{\rm hom}}^{(0)}$ the solution of
\begin{eqnarray}\label{eq:dke_order_minus2}
\nu_{zi}
\frac{T}{m_z}
\nabla_v\cdot
\left(F_{Mz}
\nabla_v
\left(
\frac{H_{b,{\rm hom}}^{(0)}}{F_{Mz}}
\right)
\right)
=
0.
\end{eqnarray}
If we multiply by $H_{b,{\rm hom}}^{(0)}/F_{Mz}$, integrate over velocities and then integrate by parts, we get
\begin{eqnarray}\label{eq:dke_order_minus2_2}
\int
\left|
\nabla_v
\left(
\frac{H_{b,{\rm hom}}^{(0)}}{F_{Mz}}
\right)
\right|^2
F_{Mz}
\dd^3v
=
0.
\end{eqnarray}
Every solution of \eq{eq:dke_order_minus2} is a solution of \eq{eq:dke_order_minus2_2}. The latter is satisfied if and only if
\begin{eqnarray}\label{eq:dke_order_minus2_3}
\nabla_v
\left(
\frac{H_{b,{\rm hom}}^{(0)}}{F_{Mz}}
\right)
=
0,
\end{eqnarray}
i.e. if and only if $H_{b,{\rm hom}}^{(0)}$ has the form
\begin{equation}\label{eq:H_2zero_hom}
H_{b,{\rm hom}}^{(0)}= \frac{N^{(0)}}{n_z} F_{Mz}.
\end{equation}
It is easy to check that \eq{eq:H_2zero_hom} satisfies \eq{eq:dke_order_minus2}. The function $N^{(0)}(\rcoor,\theta,\zeta)$ is determined by higher order equations.

\subsubsection{Particular solution of the inhomogeneous equation \eq{eq:dke_z_new_order_minus1}.}
\label{sec:solution_eq_H_2_order_minus1_inhomogeneous}

The key for the explicit calculation of $\Gamma_{z,b}$ in the Pfirsch-Schl{\"u}ter is the diagonalization of the operator ${\cal K}$. In \ref{sec:eigenfunctions_of_C_zi_diff} we find its eigenfunctions and eigenvalues, and explain how to write any function on velocity space as a linear combination of eigenfunctions of ${\cal K}$. Here, like in \ref{sec:eigenfunctions_of_C_zi_diff}, we use spherical coordinates $\{v,\beta,\phi\}$ in velocity space, where $v\in[0,\infty)$, $\beta\in[0,\pi]$ and $\phi\in[0,2\pi)$. Choose a right-handed set of orthonormal vectors $\{\eun_1, \eun_2, \bun\}$ and take a point $\bv$ of velocity space. Spherical coordinates are defined by
\begin{equation}\label{eq:spherical_coordinates}
\bv = v\cos\beta\, \bun + v\sin\beta(\cos\phi\,\eun_1 + \sin\phi\,\eun_2).
\end{equation}
In addition, some expressions are simpler in terms of
\begin{equation}
x:= \frac{m_z v^2}{2T},
\end{equation}
a variable that we will frequently use in what follows.

Noting that
\begin{eqnarray}
\fl
\matM \, : \bv\bv = {\matM}: \frac{1}{2\pi}\int_0^{2\pi}\bv\bv\dd\phi=
{\matM}: v^2\left(\cos^2\beta\bun\bun + \frac{\sin^2\beta}{2}\left(\matI - \bun\bun\right)\right),
\end{eqnarray}
denoting by $P_l$ the Legendre polynomials and by $L_p^{(\alpha)}$ the generalized Laguerre polynomials (see definitions in \ref{sec:eigenfunctions_of_C_zi_diff}), and employing $P_2(\cos\beta) = (3\cos^2\beta -1)/2$, $L_0^{(\alpha)}(x) = 1$ and $L_1^{(\alpha)}(x) = -x + \alpha + 1$, we can rewrite \eq{eq:dke_z_new_order_minus1} as
\begin{eqnarray}\label{eq:dke_z_new_order_minus1_rewritten}
\fl
{\cal K} H_b^{(0)}
=
\frac{2m_z}{T}
 \left(
 \frac{1}{3}\tr(\matM)(xP_2 + L_1^{(1/2)}) - \matM \, : \bun\bun \, x P_2
 \right)
 F_{Mz},
\end{eqnarray}
where the right-hand side is now expressed as a linear combination of eigenfunctions of ${\cal K}$ and is, therefore, immediate to solve using \eq{eq:eigenfunctions}. Then,
\begin{equation}\label{eq:H2_zero_general}
H_b^{(0)} = H_{b,{\rm hom}}^{(0)} + {\tilde H}_b^{(0)},
\end{equation}
where $H_{b,{\rm hom}}^{(0)}$ is given in \eq{eq:H_2zero_hom} and the particular solution of the inhomogeneous equation, ${\tilde H}_b^{(0)}$, is
\begin{eqnarray}\label{eq:H2_zero_tilde}
\fl
{\tilde H}_b^{(0)}
=
\frac{m_z}{T}
 \left[
   \matM \, : \bun\bun \, x P_2
 -
 \frac{1}{3}\tr(\matM)\left(xP_2 + L_1^{(1/2)}\right)
 \right]
 F_{Mz}
 .
\end{eqnarray}

\subsection{Solution of \eq{eq:dke_z_new_order_zero}}
\label{sec:solution_eq_H_2_order_zero}

Again, we need to write the right-hand side of \eq{eq:dke_z_new_order_zero} as a linear combination of eigenfunctions of the operator ${\cal K}$. Employing $P_1(\cos\beta) =\cos\beta$, $P_3(\cos\beta) = (5\cos^3\beta - 3\cos\beta)/2$, $v_{||} = \sqrt{2T/m_z} \, x^{1/2}P_1$,
\begin{equation}
P_1P_2 = \frac{3}{5} P_3 + \frac{2}{5} P_1,
\end{equation}
\begin{equation}
x^{3/2} P_1 = \frac{5}{2} x^{1/2}P_1-x^{1/2}L_1^{(3/2)}P_1,
\end{equation}
\begin{equation}
x^{3/2} P_1 P_2 = 
\frac{3}{5}x^{3/2}P_3 + x^{1/2}P_1 - \frac{2}{5}x^{1/2}L_1^{(3/2)}P_1
,
\end{equation}
\begin{equation}
\bun\cdot\nabla x = - \frac{Z_z e}{T} \bun\cdot\nabla\varphi_1,
\end{equation}
\begin{equation}
v_{||}\bun\cdot\nabla L_1^{(1/2)} = 
\sqrt{\frac{2T}{m_z}}\frac{Z_z e}{T} \bun\cdot\nabla\varphi_1 x^{1/2} P_1
\end{equation}
and
\begin{eqnarray}
\fl
v_{||}\bun\cdot\nabla (xP_2) = 
-\sqrt{\frac{2T}{m_z}}
\frac{Z_z e}{T} \bun\cdot\nabla\varphi_1 x^{1/2} P_1
\nonumber\\[5pt]
\fl
\hspace{1cm}
- 3 \bun\cdot\nabla B \sqrt{\frac{2T}{m_z}}\frac{1}{B}
\left[
-\frac{1}{5}x^{3/2} P_3 + \frac{1}{2} x^{1/2} P_1 - \frac{1}{5} x^{1/2} L_1^{(3/2)} P_1
\right],
\end{eqnarray}
we recast \eq{eq:dke_z_new_order_zero} into
\begin{eqnarray}\label{eq:dke_z_new_order_zero_more_explicit}
\fl
{\cal K} H_b^{(1)}=
\frac{1}{\nu_{zi}}\sqrt{\frac{2T}{m_z}}
\bun\cdot\nabla
\left(
\frac{N^{(0)}}{n_z}
\right)
 x^{1/2} P_1
F_{Mz}
\nonumber\\[5pt]
\fl
\hspace{0.5cm}
+
\frac{2}{\nu_{zi}}
\sqrt{\frac{m_z}{2T}}
 \Bigg\{
 - \matM \, :\bun\bun\frac{Z_z e}{T}\bun\cdot\nabla\varphi_1 x^{1/2} P_1
 \nonumber\\[5pt]
\fl
\hspace{0.5cm}
-
 \left(
\matM \, : \bun\bun 
 -
 \frac{1}{3}\tr(\matM)
 \right)
 \frac{\bun\cdot\nabla B}{B}
 \left(
 -\frac{3}{5}x^{3/2}P_3 + \frac{3}{2}x^{1/2} P_1 - \frac{3}{5}x^{1/2}L_1^{(3/2)} P_1
 \right)
\nonumber\\[5pt]
\fl
\hspace{0.5cm}
 +
 \left(
 \bun\cdot\nabla(\matM \, : \bun\bun) 
 -
 \frac{1}{3}\bun\cdot\nabla\tr(\matM)
 \right)\Big(\frac{3}{5}x^{3/2}P_3 + x^{1/2}P_1
 - \frac{2}{5}x^{1/2}L_1^{(3/2)}P_1\Big)
 \nonumber\\[5pt]
\fl
\hspace{0.5cm}
 -\frac{1}{3}
\bun\cdot\nabla
\tr(\matM)
x^{1/2}P_1
\left(L_1^{(3/2)} - 1\right)
 \Bigg\}F_{Mz}
.
\end{eqnarray}
The solution of the homogeneous equation is irrelevant for us and we can set it to zero. The solution of the inhomogeneous equation is
\begin{eqnarray}\label{eq:H2_one}
\fl
H_b^{(1)}=
-\frac{1}{\nu_{zi}}\sqrt{\frac{2T}{m_z}}
\bun\cdot\nabla
\left(
\frac{N^{(0)}}{n_z}
\right)
 x^{1/2} P_1
F_{Mz}
\nonumber\\[5pt]
\fl
\hspace{0.5cm}
+
\frac{2}{\nu_{zi}}
\sqrt{\frac{m_z}{2T}}
 \Bigg\{
  \matM:\bun\bun\frac{Z_z e}{T}\bun\cdot\nabla\varphi_1 x^{1/2} P_1
 \nonumber\\[5pt]
\nonumber\\[5pt]
\fl
\hspace{0.5cm}
-
 \left(
  \matM \, : \bun\bun 
 -
 \frac{1}{3}\tr(\matM)
 \right)
 \frac{\bun\cdot\nabla B}{B}
 \left(
 \frac{1}{5}x^{3/2}P_3 - \frac{3}{2}x^{1/2} P_1 + \frac{1}{5}x^{1/2}L_1^{(3/2)} P_1
 \right)
\nonumber\\[5pt]
\fl
\hspace{0.5cm}
 +
 \left(
 \bun\cdot\nabla(\matM \, : \bun\bun) 
 -
 \frac{1}{3}\bun\cdot\nabla\tr(\matM)
 \right)\Big(-\frac{1}{5}x^{3/2}P_3 - x^{1/2}P_1
 + \frac{2}{15}x^{1/2}L_1^{(3/2)}P_1\Big)
 \nonumber\\[5pt]
\fl
\hspace{0.5cm}
 -\frac{1}{3}
\bun\cdot\nabla
\tr(\matM)
x^{1/2}P_1
\left(-\frac{1}{3}L_1^{(3/2)} + 1\right)
 \Bigg\}F_{Mz}
.
\end{eqnarray}

\subsection{Solvability condition \eq{eq:H2_solvability_condition}}
\label{sec:solution_eq_H_2_solvability_condition}

Velocity space integrals in coordinates $x = m_z v^2 /(2T)$ and $\xi := \cos\beta = v_{||} / v$ read
\begin{equation}\label{eq:velocity_integrals_x_beta}
\int (\cdot) \, \dd^3v = \pi \left(\frac{2T}{m_z}\right)^{3/2} \int_0^\infty x^{1/2}\dd x \int_{-1}^1 \dd \xi \,(\cdot).
\end{equation}
Applying this to the left side of \eq{eq:H2_solvability_condition}, and employing the orthogonality relations of the Legrendre and Laguerre polynomials (see \eq{eq:orthogonality_relations_Legendre} and \eq{eq:orthogonality_relations_Laguerre}), we get
\begin{eqnarray}\label{eq:left_side_H2_solvability_condition}
\fl
\frac{1}{\nu_{zi}}\frac{n_z}{B}
\Bigg\{
-\frac{T}{m_z}
\bun\cdot\nabla
\left(
\frac{N^{(0)}}{n_z}
\right)
+
  \matM \,:\bun\bun \frac{Z_z e}{T}\bun\cdot\nabla\varphi_1
\nonumber\\[5pt]
\fl
\hspace{0.5cm}
+\frac{3}{2}
 \left(
   \matM \, : \bun\bun 
 -
 \frac{1}{3}\tr(\matM)
 \right)
 \frac{\bun\cdot\nabla B}{B}
 -
  \bun\cdot\nabla(\matM \, : \bun\bun) 
 \Bigg\}
 =
 Q(r)
.
\end{eqnarray}
Multiplying this equation by $B^2 / n_z$ and flux-surface averaging, we find
\begin{eqnarray}\label{eq:Q_solvability_condition}
\fl
 Q(r)
 =
  \frac{1}{\nu_{zi}}
 \left\langle
 \frac{B^2}{n_z}
 \right\rangle^{-1}
 \Bigg\langle
  \matM \, : \bun\bun 
 \frac{Z_z e}{T}\bB\cdot\nabla\varphi_1
 +
 \frac{3}{2}
  \left(
   \matM \, : \bun\bun 
 -
 \frac{1}{3}\tr(\matM)
 \right) 
 \bun\cdot\nabla B
 \Bigg\rangle
.
\end{eqnarray}

Finally, we rewrite \eq{eq:left_side_H2_solvability_condition} as
\begin{eqnarray}\label{eq:left_side_H2_solvability_condition_final}
\fl
\bB\cdot\nabla
\left(
\frac{N^{(0)}}{n_z}
 +
 \frac{m_z}{T} \matM \, : \bun\bun
 \right) =
 \nonumber\\[5pt]
\fl
\hspace{0.5cm}
\frac{m_z}{T}
\Bigg[
  \matM \, : \bun\bun 
 \frac{Z_z e}{T}\bB\cdot\nabla\varphi_1
 +\frac{3}{2}\left(
  \matM \, : \bun\bun 
 -
 \frac{1}{3}\tr(\matM)
 \right)
 \bun\cdot\nabla B
 \nonumber\\[5pt]
\fl
\hspace{0.5cm}
 -
\frac{B^2}{n_z}
 \left\langle
 \frac{B^2}{n_z}
 \right\rangle^{-1}
 \Bigg\langle
 \matM \, : \bun\bun 
 \frac{Z_z e}{T}\bB\cdot\nabla\varphi_1
 +
 \frac{3}{2}
  \left(
  \matM \, : \bun\bun 
 -
 \frac{1}{3}\tr(\matM)
 \right) 
 \bun\cdot\nabla B
 \Bigg\rangle
 \Bigg]
.
\end{eqnarray}
This is a magnetic differential equation that can be solved, numerically, determining $N^{(0)}/n_z$ up to an irrelevant additive flux function. For $\epsilon \ll 1$, the equation can be solved analytically. This is due to the fact that the right-hand side can be dropped because it is a factor of $\epsilon$ smaller than the second term on the left-hand side. Hence, we can approximate \eq{eq:left_side_H2_solvability_condition_final} by
\begin{eqnarray}\label{eq:left_side_H2_solvability_condition_final_final}
\fl
\bB\cdot\nabla
\left(
\frac{N^{(0)}}{n_z}
 +
 \frac{m_z}{T} \matM \, : \bun\bun
 \right) = 0,
\end{eqnarray}
giving
\begin{eqnarray}\label{eq:Nzero_over_nz}
\frac{N^{(0)}}{n_z} = 
 -
 \frac{m_z}{T} \matM \, : \bun\bun
 ,
\end{eqnarray}
up to an additive flux function. Recalling \eq{eq:Gammab_PS}, \eq{eq:H_2zero_hom}, \eq{eq:H2_zero_general} and \eq{eq:H2_zero_tilde}, we compute $\Gamma_{z,b}$ in subsection \ref{sec:explicit_expression_Gammab_PS}.

\subsection{Explicit expression for $\Gamma_{z,b}$}
\label{sec:explicit_expression_Gammab_PS}

With the results of previous subsections, we can calculate the right-hand side of \eq{eq:Gammab_PS}. To that end, it is convenient to use coordinates $x$ and $\beta$, and make repeated use of expansions in terms of Legendre and generalized Laguerre polynomials (see \ref{sec:eigenfunctions_of_C_zi_diff}). First, note that
\begin{equation}
\fl
\bv_d\cdot\nabla r
= \frac{1}{B}(\bun\times\nabla\varphi_1)\cdot\nabla r
+
\frac{2T}{m_z}\frac{1}{B\Omega_z}
\left(
-\frac{2}{3}L_1^{(1/2)} + 1 + \frac{1}{3}xP_2
\right) (\bun\times\nabla B)\cdot \nabla r,
\end{equation}
where we have used that, in a magnetohydrodynamic equilibrium $(\nabla\times\bB)\cdot\nabla r = 0$, so that $\bun\times(\bun\cdot\nabla\bun) = B^{-1}\bun\times\nabla B$. Then, we can write
\begin{eqnarray}\label{eq:Gammab_I_1_I_2}
\Gamma_{z,b}
=
\left\langle
\frac{1}{B}(\bun\times\nabla\varphi_1)\cdot\nabla r I_1
\right\rangle
+
\left\langle
\frac{2T}{m_z}\frac{1}{B\Omega_z}
 (\bun\times\nabla B)\cdot \nabla r I_2
\right\rangle,
\end{eqnarray}
where
\begin{eqnarray}\label{eq:I1}
\fl
I_1 = \int H_{b, {\rm hom}}^{(0)} \dd^3 v + \int \tilde H_b^{(0)} \dd^3 v
\end{eqnarray}
and
\begin{eqnarray}\label{eq:I2}
\fl
I_2 = \int  \left(
-\frac{2}{3}L_1^{(1/2)} + 1 + \frac{1}{3}xP_2
\right) H_{b, {\rm hom}}^{(0)} \dd^3 v
\nonumber\\[5pt]
\fl\hspace{1cm}
+
\int  \left(
-\frac{2}{3}L_1^{(1/2)} + 1 + \frac{1}{3}xP_2
\right) \tilde H_b^{(0)} \dd^3 v.
\end{eqnarray}

We start by computing $I_1$. It is immediate to find that
\begin{eqnarray}
 \int H_{b, {\rm hom}}^{(0)} \dd^3 v
 =
 -\frac{n_z m_z}{T}{\matM} : \bun\bun.
\end{eqnarray}
For the second term on the right-hand side of \eq{eq:I1}, one gets
\begin{eqnarray}
\int \tilde H_b^{(0)} \dd^3 v = 0,
\end{eqnarray}
which is easy to obtain using \eq{eq:velocity_integrals_x_beta} and \eq{eq:H2_zero_tilde}. Hence,
\begin{eqnarray}\label{eq:result_I1}
 I_1
 =
 -\frac{n_z m_z}{T}{\matM} : \bun\bun.
\end{eqnarray}

As for $I_2$, using
\eq{eq:H_2zero_hom}, \eq{eq:Nzero_over_nz} and \eq{eq:velocity_integrals_x_beta}, we get
\begin{eqnarray}\label{eq:result_I2_partial1}
\fl
\int \left(
-\frac{2}{3}L_1^{(1/2)} + 1 + \frac{1}{3}xP_2
\right) H_{b, {\rm hom}}^{(0)} \dd^3 v
= -\frac{n_z m_z}{T}{\matM} : \bun\bun,
\end{eqnarray}
where the orthogonality relations \eq{eq:orthogonality_relations_Legendre} and \eq{eq:orthogonality_relations_Laguerre} have been employed in the last step. Analogous, although a bit lengthier, manipulations allow us to calculate the second term on the right-hand side of \eq{eq:I2},
\begin{eqnarray}\label{eq:result_I2_partial2}
\fl
\int \left(
-\frac{2}{3}L_1^{(1/2)} + 1 + \frac{1}{3}xP_2
\right) \tilde H_b^{(0)} \dd^3 v
=
\frac{n_z m_z}{4T}\left({\matM} : \bun\bun + \tr(\matM)\right)
.
\end{eqnarray}
Then, from \eq{eq:result_I2_partial1} and \eq{eq:result_I2_partial2},
\begin{eqnarray}
I_2
=
\frac{n_z m_z}{4T}\left(
\tr(\matM)-
3{\matM} : \bun\bun\right)
.
\end{eqnarray}
Finally, for $\epsilon \ll 1$,
\begin{eqnarray}\label{eq:Gammab_PS_explicit_result}
\fl
\Gamma_{z,b}
=
-\frac{m_z\eta}{B_0 T}
\left\langle
 {\matM} : \bun\bun \,
(\bun\times\nabla\varphi_1)\cdot\nabla r 
\right\rangle
\nonumber\\[5pt]
\fl
\hspace{1cm}
+
\frac{m_z\eta}{2 Z_z e B_0^2}
\left\langle
\left(
\tr(\matM)-
3{\matM} : \bun\bun\right)
 (\bun\times\nabla B)\cdot \nabla r
\right\rangle,
\end{eqnarray}
where $\eta$ has been defined in \eq{eq:def_n_z} and $B_0$ has been defined in \eq{eq:ordering_B}.

\subsection{Comparison of the relative sizes of $\Gamma_{z,a}$ and $\Gamma_{z,b}$}
\label{sec:comparison_Gammaa_vs_Gammab_PS}

Recalling \eq{eq:sizeM} and \eq{eq:Gammab_PS_explicit_result}, we find
\begin{equation}
\Gamma_{z,b} 
\sim \epsilon^{1/2} \rho_{z*}  \frac{h_{i+}}{F_{Mi}} n_z v_{tz}.
\end{equation}
And, if the main ions are in the $1/\nu$ regime (see \eq{eq:hi_over_FMi_1overnu}),
\begin{equation}\label{eq:size_Gammab_PS}
\Gamma_{z,b} 
\sim \epsilon^{1/2} \rho_{z*}  \frac{\rho_{i*}}{\nu_{ii*}} n_z v_{tz}.
\end{equation}
It is convenient to employ the relations
\begin{equation}\label{eq:relation_rhoz_rhoi}
\rho_{z*} \sim \frac{1}{Z_z}\sqrt{\frac{m_z}{m_i}}\rho_{i*}
\end{equation}
and
\begin{equation}\label{eq:relation_nuzi_nuii}
\nu_{zi*} \sim Z_z^2 \sqrt{\frac{m_i}{m_z}} \nu_{ii*}
\end{equation}
to write the estimate \eq{eq:size_Gammab_PS} in terms of $\rho_{z*}$ and $\nu_{zi*}$. Then,
\begin{equation}\label{eq:size_Gammab_PS_final}
\Gamma_{z,b} 
\sim \epsilon^{1/2} Z_z^3 \frac{m_i}{m_z}\frac{\rho_{z*}^2}{\nu_{zi*}} n_z v_{tz}.
\end{equation}

Let us rewrite the estimate \eq{eq:Gammaa_PS} for the size of $\Gamma_{z,a}$ in a slightly different fashion,
\begin{equation}\label{eq:Gammaa_PS_again}
\Gamma_{z,a} \sim  \epsilon^{-1} Z_z \rho_{z*}^2 \nu_{zi*} n_z v_{tz}
.
\end{equation}
Taking the quotient of \eq{eq:size_Gammab_PS_final} and \eq{eq:Gammaa_PS_again} it is easy to find the condition for the main ion pressure anisotropy to matter, assuming that the main ions are in the $1/\nu$ regime and the impurities are in the Pfirsch-Schl{\"u}ter regime. Since
\begin{equation}
\frac{\Gamma_{z,b}}{\Gamma_{z,a}} \sim \epsilon^{3/2} Z_z^2 \frac{m_i}{m_z}\frac{1}{\nu_{zi*}^2},
\end{equation}
the condition is
\begin{equation}
\frac{\aleph}{\nu_{zi*}^2} \gtrsim 1,
\end{equation}
where
\begin{equation}\label{eq:def_tau}
\aleph := \epsilon^{3/2} Z_z^2 \frac{m_i}{m_z}
\end{equation}
only depends on the magnetic geometry (via $\epsilon$), on the impurity charge and on the ratio of main ion to impurity mass. The parameter $\aleph$ will repeatedly appear when we evaluate the relative weight of $\Gamma_{z,a}$ and $\Gamma_{z,b}$ in different collisionality regimes.

The conditions $\nu_{ii*}\ll\epsilon^{3/2}$, $\nu_{zi*} \gg 1$ and $\aleph/\nu_{zi*}^2 \gg 1$ can be simultaneously satisfied for highly-charged impurities, although the region of parameter space in which this happens is not large. This can be seen by expressing condition $\aleph/\nu_{zi*}^2 \gg 1$ as
\begin{equation}\label{eq:cond1}
Z_z^2 \nu_{ii*} \ll \frac{\epsilon^{3/2}}{\nu_{ii*}} \, 
\end{equation}
and noting that $\nu_{zi*} \gg 1$ implies
\begin{equation}\label{eq:cond2}
Z_z^2 \nu_{ii*} \gg \sqrt{\frac{m_z}{m_i}} \, .
\end{equation}

\section{Impurities in the plateau regime}
\label{sec:impurities_plateau_regime}

We deal with the regime
\begin{equation} \label{eq:def_plateau_regime}
\epsilon^{3/2} \ll \nu_{zi*} \ll 1,
\end{equation}
in which passing particles are collisionless and trapped particles are collisional. This can be seen by comparing the parallel streaming and collision terms in \eq{eq:DKE_impurities_more_explicit}. For passing particles, $v_{||}\sim v_{tz}$ and the effective collision frequency, $\nu_{zi, {\rm eff}} \sim v_{tz}^2 \nu_{zi}/v_{||}^2 \sim \nu_{zi}$, so the collision term is negligible compared to the parallel streaming term in the regime defined by \eq{eq:def_plateau_regime}. However, for trapped particles, $v_{||}\sim \epsilon^{1/2} v_{tz}$ and $\nu_{zi, {\rm eff}} \sim v_{tz}^2 \nu_{zi}/v_{||}^2 \sim \nu_{zi}/\epsilon$, implying that, if \eq{eq:def_plateau_regime} is satisfied, the collision term is larger than the parallel streaming term. Although not in the context of impurity transport in stellarators but in the context of plasma transport in tokamaks, a good treatment of the plateau regime can be found in reference \cite{Helander_book_2002}.

Trapped particles, being collisional, give negligible transport, and we focus on passing particles. It is useful to write the drift-kinetic equation in terms of
\begin{equation}
G_z = F_{z1} - \frac{m_z A v_{||}}{T} F_{Mz},
 \end{equation}
so that \eq{eq:DKE_impurities_more_explicit} is recast into
\begin{eqnarray}\label{eq:dke_z_plateau_0}
v_{||}\bun\cdot\nabla
G_z -  \nu_{zi}
{\cal K} G_z
= S,
\end{eqnarray}
with
\begin{eqnarray}\label{eq:def_S}
\fl
S = -
 \bv_d \cdot\nabla r \Upsilon F_{Mz}
 - v_{||}\bun\cdot\nabla
\left(
 \frac{m_z A v_{||}}{T}
F_{Mz}
\right)
\nonumber\\[5pt]
\fl\hspace{1cm}
+\nu_{zi}\frac{m_z}{T}
\left(
-\tr(\matM)  + \frac{m_z}{T} {\matM} :
\left[
v_{||}^2\bun\bun + \frac{v_\perp^2}{2}\left(\matI - \bun\bun\right)
\right]
\right) F_{Mz}.
\end{eqnarray}

For passing particles, we can define the transit average of a function $f(r,\theta,\zeta,\cE,\mu)$ as
\begin{equation}
\overline{f}(r,\cE,\mu) =
\left\langle
\frac{B}{v_{||}}f
\right\rangle.
\end{equation}
It is convenient to split \eq{eq:dke_z_plateau_0} into two equations, one whose source is $\overline{S}$ and another one whose source is $S-\overline{S}$. Hence, we take $G_z = g_z + G_{z,p}$, where $g_z$ and $G_{z,p}$ are the solutions of
\begin{eqnarray}\label{eq:dke_z_average}
v_{||}\bun\cdot\nabla
g_z -  \nu_{zi}
{\cal K} g_z
= \overline{S}
\end{eqnarray}
and
\begin{eqnarray}\label{eq:dke_z_fluctuating}
v_{||}\bun\cdot\nabla
G_{z,p} -  \nu_{zi}
{\cal K} G_{z,p}
= S - \overline{S}.
\end{eqnarray}
The transit average of $S$ is
\begin{equation}
\overline{S}
=
\nu_{zi}\frac{m_z}{T}
\overline{
\left(
-\tr(\matM)  + \frac{m_z}{T} {\matM} :
\left[
v_{||}^2\bun\bun + \frac{v_\perp^2}{2}\left(\matI - \bun\bun\right)
\right]
\right)
} \, F_{Mz}.
\end{equation}
The subindex $p$ in $G_{z,p}$ indicates that this piece of the impurity distribution gives the plateau contribution, as we show in subsection \ref{sec:impurity_flux_due_to_Gz2}.

\subsection{Impurity flux due to $g_{z}$}
\label{sec:impurity_flux_due_to_Gz1}

Radial impurity transport due to $g_{z}$ is small. It is not difficult to show this by expanding $g_{z} = g_{z}^{(0)} +  g_{z}^{(1)} + g_{z}^{(2)} + \dots$ in powers of $\nu_{zi*} \ll 1$. To lowest order in this expansion, one finds
\begin{equation}
v_{||}\bun\cdot\nabla
g_{z}^{(0)} = 0,
\end{equation}
implying that $g_{z}^{(0)}$ is a flux function, which, as a consequence, does not give transport. To next order, we have the equation
\begin{eqnarray}\label{eq:equation_Gz10}
v_{||}\bun\cdot\nabla
g_{z}^{(1)} -  \nu_{zi}
{\cal K} g_{z}^{(0)}
= \overline{S}.
\end{eqnarray}
Its transit average,
\begin{eqnarray}\label{eq:equation_Gz10_averaged}
 -  \nu_{zi}
\overline{{\cal K} g_{z}^{(0)}}
= \overline{S},
\end{eqnarray}
determines $g_{z}^{(0)} \sim {\nu_{zi}^{-1}} \overline{S}$. Subtracting \eq{eq:equation_Gz10_averaged} to \eq{eq:equation_Gz10}, one finds
\begin{eqnarray}\label{eq:equation_Gz11}
 v_{||}\bun\cdot\nabla
g_{z}^{(1)}
+ \nu_{zi}
\left(
\overline{{\cal K} g_{z}^{(0)}}
 -  
{\cal K} g_{z}^{(0)}
\right)
= 0.
\end{eqnarray}
From this equation, we deduce that $g_{z}^{(1)} \sim \nu_{zi_*} g_{z}^{(0)}$. Note that $g_{z}^{(1)}$ is odd in $v_{||}$ (because $g_{z}^{(0)}$ is even) and then $\Gamma_z[g_{z}^{(1)}] = 0$. The lowest-order piece of the expansion contributing to $\Gamma_z$ is $g_{z}^{(2)} \sim \nu_{zi_*}^2 g_{z}^{(0)} \sim \epsilon^{1/2} \nu_{zi_*}^2  (h_{i+}/F_{Mi}) F_{Mz}$. If the main ions are in the $1/\nu$ regime, this implies that
\begin{equation}\label{eq:Gammaz_Gz12}
\Gamma_z[g_{z}^{(2)}]
\sim \frac{\epsilon^{1/2} Z_z^3 m_i}{m_z} \nu_{zi_*}\rho_{z*}^2 n_z v_{tz}.
\end{equation}
We will see that this contribution is negligible compared to the one due to $G_{z,p}$.

\subsection{Impurity flux due to $G_{z,p}$}
\label{sec:impurity_flux_due_to_Gz2}

\subsubsection{Solution of equation \eq{eq:dke_z_fluctuating}.}
\label{sec:solution_dke_Gz2}

The impurity flux due to $G_{z,p}$ is produced by a small collisional layer around $v_{||} = 0$. In order to treat \eq{eq:dke_z_fluctuating}, we will employ coordinates $u$ and $\mu$, where now we are denoting the parallel velocity by $u$ (we have changed the notation for the parallel velocity from $v_{||}$ to $u$ to stress that now it is viewed as an independent variable). The total energy per unit mass reads, in terms of these coordinates,
\begin{equation}
\cE = \frac{1}{2}u^2 + \mu B + \frac{Z_z e (\varphi_0 + \varphi_1)}{m_z}.
\end{equation}
Denoting by $\hat G_{z,p}$ the function $G_{z,p}$ expressed in coordinates $u$ and $\mu$, \eq{eq:dke_z_fluctuating} becomes
\begin{eqnarray}\label{eq:dke_z_plateau}
\fl
\Bigg[
u\bun\cdot\nabla - 
\left(
\mu \bun\cdot\nabla B + \frac{Z_z e}{m_z}\bun\cdot\nabla\varphi_1
\right)\partial_u
\Bigg]
\hat G_{z,p}
-
\nu_{zi}
{\cal K} \hat G_{z,p}
=
\tilde{S},
\end{eqnarray}
where $\tilde S = S - \overline{S}$ is assumed to be expressed in terms of $u$ and $\mu$.

We expect $\hat G_{z,p}$ to have large derivatives with respect to $u$ in the small collisional layer of interest. Then, we can take
\begin{equation}
\nu_{zi}{\cal K} \hat G_{z,p} \simeq \nu_{zi}\frac{T}{m_z}\partial_u^2\hat G_{z,p}.
\end{equation}
By balancing
\begin{equation}
u\bun\cdot\nabla\hat G_{z,p}
\sim
\nu_{zi}\frac{T}{m_z}\partial_u^2\hat G_{z,p},
\end{equation}
we find the size of the layer in the coordinate $u$, $\Delta_u$,
\begin{equation}\label{eq:size_plateau_layer}
\frac{\Delta_u}{v_{tz}} \sim \nu_{zi*}^{1/3}.
\end{equation}
Then, using $\epsilon^{3/2}\ll \nu_{zi*}$ on the parallel streaming term on the left-hand side of \eq{eq:dke_z_plateau}, we find
\begin{equation}
\left|\left(
\mu \bun\cdot\nabla B + \frac{Z_z e}{m_z}\bun\cdot\nabla\varphi_1
\right)\partial_u\hat G_{z,p}
\right|
\ll
|u\bun\cdot\nabla \hat G_{z,p}|.
\end{equation}

Next, we simplify $\tilde S = S - \overline{S}$ (recall \eq{eq:def_S}). First, employing $\Delta_u / v_{tz} \sim \nu_{zi*}^{1/3} \ll 1$, we get
\begin{equation}
|u^2\bun\cdot\nabla A|
\ll
\left|\left(
\mu \bun\cdot\nabla B_1 + \frac{Z_z e}{m_z}\bun\cdot\nabla\varphi_1
\right) A_0
\right|,
\end{equation}
where $A_0$ has been introduced in \eq{eq:A_for_small_epsilon}. An explicit expression for $A_0$ can be found below, in equation \eq{eq:explicit_expression_A0}.

Second, expanding in $\epsilon$ and noting that in the layer $v_{||}^2 \ll v_\perp^2/2$, we have
\begin{eqnarray}
\fl
\nu_{zi}\frac{m_z}{T}
\left(
-\tr(\matM)  + \frac{m_z}{T} {\matM} :
\left[
v_{||}^2\bun\bun + \frac{v_\perp^2}{2}\left(\matI - \bun\bun\right)
\right]
\right)
 \, F_{Mz}
 \nonumber\\[5pt]
 \fl\hspace{1cm}
-
\nu_{zi}\frac{m_z}{T}
\overline{
\left(
-\tr(\matM)  + \frac{m_z}{T} {\matM} :
\left[
v_{||}^2\bun\bun + \frac{v_\perp^2}{2}\left(\matI - \bun\bun\right)
\right]
\right)
} \, F_{Mz}
\simeq
 \nonumber\\[5pt]
 \fl\hspace{1cm}
\nu_{zi}\frac{m_z}{T}
\left(
-\tr(\matM)  + \mu B_0 \frac{m_z}{T} {\matM} :
  \left(\matI - \bun\bun\right)
\right)
 \, \hat F_{Mz}
  \nonumber\\[5pt]
 \fl\hspace{1cm}
+
\nu_{zi}\frac{m_z}{T}
\left\langle
\tr(\matM)  - \mu B_0 \frac{m_z}{T} {\matM} :
  \left(\matI - \bun\bun\right)
\right\rangle
 \, \hat F_{Mz},
\end{eqnarray}
where
\begin{equation}\label{eq:F_M_simplified}
\hat F_{Mz} = \eta
\left(
\frac{m_z}{2\pi T}
\right)^{3/2} \exp\left(-\frac{m_z \mu B_0}{T}\right).
\end{equation}
Recall that, to lowest order in $\epsilon$,
\begin{equation}
\eta(r) \simeq n_z,
\end{equation}
and the impurity density is a flux function.

We can rewrite \eq{eq:dke_z_plateau} as
\begin{eqnarray}\label{eq:dke_z_plateau_2}
\fl
u\bun\cdot\nabla
\hat G_{z,p}
-
\nu_{zi}\frac{T}{m_z}\partial_u^2\hat G_{z,p}
= 
-
 \bv_d \cdot\nabla r \hat \Upsilon \hat F_{Mz}
+
 \frac{m_z A_0}{T}
\left(
\mu \bun\cdot\nabla B_1 + \frac{Z_z e}{m_z}\bun\cdot\nabla\varphi_1
\right)
\hat F_{Mz}
 \nonumber\\[5pt]
\fl\hspace{1cm}
 - \nu_{zi}\frac{m_z}{T}
\left(
\tr(\matM) - \langle \tr(\matM) \rangle 
\right)
 \, \hat F_{Mz}
 \nonumber\\[5pt]
\fl\hspace{1cm}
 +
 \nu_{zi}\frac{m_z^2}{T^2}\mu B_0
\left(
   {\matM} :
  \left(\matI - \bun\bun\right)
  -\left\langle
   {\matM} :
  \left(\matI - \bun\bun\right)
  \right\rangle
\right)
 \, \hat F_{Mz}
,
\end{eqnarray}
with
\begin{equation}\label{eq:Upsilon_simplified}
\hat\Upsilon \simeq \frac{\eta'}{\eta} + \frac{Z_ze\varphi'_0}{T}
+\frac{T'}{T}\left(
\frac{m_z\mu B_0}{T} - \frac{3}{2}
\right).
\end{equation}

The two first terms on the right-hand side of \eq{eq:dke_z_plateau_2} have a size $O(a^{-1}\rho_{i*} v_{ti} \hat F_{Mz})$ and the two last ones have a size $O(\nu_{zi} \epsilon^{1/2} (h_{i+}/F_{Mi})\hat F_{Mz})$.

In order to give a completely explicit calculation, we will be more specific about our spatial coordinates. We use Boozer coordinates~\cite{Boozer81}, so that the magnetic field can be simultaneously written as
\begin{equation}\label{eq:B_contrav}
\bB = \Psi'_t(\rcoor)\nabla\rcoor\times\nabla(\theta - \iota(r)\zeta)
\end{equation}
and
\begin{equation}\label{eq:B_cov}
\bB = \beta\nabla r + I_t(r) \nabla\theta + I_p(r) \nabla\zeta,
\end{equation}
where $\iota(\rcoor)$ is the rotational transform, $\Psi_t(\rcoor)$ is the toroidal magnetic flux over $2\pi$, and $I_t$ and $I_p$ are the toroidal and poloidal currents, respectively. In Boozer coordinates, the volume element is conveniently written in terms of the magnitude of $\bB$,
\begin{equation}\label{eq:volumeelement_Boozer}
\sqrt{g_{V}} = \frac{V'(r)}{4\pi^2}\frac{\langle B^2 \rangle}{B^2}.
\end{equation}
Note that, to lowest order in $\epsilon$,
\begin{equation}
\sqrt{g_{V}} \simeq V'(r)/(4\pi^2) \sim \epsilon R_0^2.
\end{equation}

The parallel streaming operator in Boozer coordinates is
\begin{equation}
\bun\cdot\nabla = 
\frac{\Psi'_t}{B}\frac{1}{\sqrt{g_{V}}}(\partial_\zeta + \iota \partial_\theta)
\end{equation}
and to lowest order in $\epsilon$,
\begin{equation}
\bun\cdot\nabla \simeq 
\frac{\Psi'_t}{B_0}\frac{4\pi^2}{V'}(\partial_\zeta + \iota \partial_\theta).
\end{equation}

In Boozer coordinates, the radial drift reads
\begin{eqnarray}
\fl
\bv_d\cdot\nabla\rcoor
=
\frac{u^2 + 2\mu B}{\Omega_z B^2}\left(
\frac{\mu B}{u^2 + 2\mu B} - 1
\right)\frac{1}{\sqrt{g_{V}}}\left(
I_p \partial_\theta B - I_t \partial_\zeta B
\right)
\nonumber\\[5pt]
\fl\hspace{1cm}
+
\frac{I_t \partial_\zeta\varphi_1 - I_p \partial_\theta\varphi_1}{B^2\sqrt{g_{V}}}
.
\end{eqnarray}
Taking an $\epsilon \ll 1$ expansion and using that in the layer $u^2 \ll \mu B $, we have
\begin{eqnarray}\label{eq:v_d_simplified}
\fl
\bv_d\cdot\nabla\rcoor
\simeq
-\frac{4\pi^2 m_z \mu}{Z_z e B_0^2V'}
\left(
I_p \partial_\theta B_1 - I_t \partial_\zeta B_1
\right)
\nonumber\\[5pt]
\fl\hspace{1cm}
+\frac{4\pi^2}{B_0^2V'}
(I_t \partial_\zeta\varphi_1 - I_p \partial_\theta\varphi_1)
.
\end{eqnarray}

Equation \eq{eq:dke_z_plateau_2} becomes
\begin{eqnarray}\label{eq:dke_z_plateau_3}
u \frac{\Psi'_t}{B_0}\frac{4\pi^2}{V'}
\left(
\partial_\zeta + \iota\partial_\theta
\right)
\hat G_{z,p}
-
\nu_{zi}\frac{T}{m_z}\partial_u^2\hat G_{z,p}
=
\tilde S,
\end{eqnarray}
with
\begin{eqnarray}\label{eq:source_layer_plateau}
\fl\tilde S
\simeq
-
 \bv_d \cdot\nabla r \hat \Upsilon \hat F_{Mz}
+
 \frac{m_z A_0}{T}
\left(
\frac{v^2}{2B_0} \bun\cdot\nabla B_1 + \frac{Z_z e}{m_z}\bun\cdot\nabla\varphi_1
\right)
\hat F_{Mz}
 \nonumber\\[5pt]
\fl\hspace{1cm}
 - \nu_{zi}\frac{m_z}{T}
\left(
\tr(\matM) - \langle \tr(\matM) \rangle 
\right)
 \, \hat F_{Mz}
 \nonumber\\[5pt]
\fl\hspace{1cm}
 +
 \nu_{zi}\frac{m_z^2}{T^2}\frac{v^2}{2}
\left(
   {\matM} :
  \left(\matI - \bun\bun\right)
  -\left\langle
   {\matM} :
  \left(\matI - \bun\bun\right)
  \right\rangle
\right)
 \, \hat F_{Mz}
.
\end{eqnarray}
In \eq{eq:source_layer_plateau}, $\hat F_{Mz}$, $\hat\Upsilon$ and $\bv_d\cdot\nabla r$ must be understood as given by \eq{eq:F_M_simplified}, \eq{eq:Upsilon_simplified} and \eq{eq:v_d_simplified}.

At this point, it is useful to split \eq{eq:dke_z_plateau_3} into two equations, in analogy to the treatment of Section \ref{sec:impurities_PS_regime}. We take
\begin{equation}\label{eq:decomposition_Ha_Hb_plateau}
\hat G_{z,p} = H_a + H_b,
\end{equation}
where $H_a$ and $H_b$ are the solutions of 
\begin{eqnarray}\label{eq:dke_z_plateau_3_a}
u\frac{\Psi'_t}{B_0}\frac{4\pi^2}{V'}
\left(
\partial_\zeta + \iota\partial_\theta
\right)
H_a
-
\nu_{zi}\frac{T}{m_z}\partial_u^2 H_a
=
\tilde S_a
\end{eqnarray}
and
\begin{eqnarray}\label{eq:dke_z_plateau_3_b}
u\frac{\Psi'_t}{B_0}\frac{4\pi^2}{V'}
\left(
\partial_\zeta + \iota\partial_\theta
\right)
H_b
-
\nu_{zi}\frac{T}{m_z}\partial_u^2 H_b
=
\tilde S_b,
\end{eqnarray}
where
\begin{eqnarray}\label{eq:source_layer_plateau_a}
\fl\tilde S_a
=
-
 \bv_d \cdot\nabla r \hat \Upsilon \hat F_{Mz}
+
 \frac{m_z A_0}{T}
\left(
\mu \bun\cdot\nabla B_1 + \frac{Z_z e}{m_z}\bun\cdot\nabla\varphi_1
\right)
\hat F_{Mz}
\end{eqnarray}
and
\begin{eqnarray}\label{eq:source_layer_plateau_b}
\fl\tilde S_b
=
 - \nu_{zi}\frac{m_z}{T}
\left(
\tr(\matM) - \langle \tr(\matM) \rangle 
\right)
 \, \hat F_{Mz}
 \nonumber\\[5pt]
\fl\hspace{1cm}
 +
 \nu_{zi}\frac{m_z^2}{T^2}\mu B_0
\left(
  {\matM} :
  \left(\matI - \bun\bun\right)
  -\left\langle
   {\matM} :
  \left(\matI - \bun\bun\right)
  \right\rangle
\right)
 \, \hat F_{Mz}
.
\end{eqnarray}
It is convenient to write explicitly $A_0$, that can be read off from equation (24) of reference \cite{Calvo2018b}. Namely,
\begin{equation}\label{eq:explicit_expression_A0}
A_0 = \frac{f_s + \langle B^2 u \rangle}{1-f_c}\frac{T}{Z_i e B_0}
\left(
\frac{n'_i}{n_i} + \frac{Z_i e \varphi'_0}{T} - 0.17 \frac{T'}{T}
\right).
\end{equation}
Here, $u$ is the solution of $\bun\cdot\nabla u = 2 B^{-3}(\bun\times\nabla r)\cdot\nabla B$ with vanishing boundary condition for $u$ at the point of the flux surface where $B$ takes the value $B_{\rm max}$, and the flux functions $f_c$ and $f_s$ are given in \ref{sec:f_c_and_f_s}.

Equations \eq{eq:dke_z_plateau_3_a} and \eq{eq:dke_z_plateau_3_b} are solved in Fourier space. We write
\begin{eqnarray}\label{eq:dke_z_plateau_3_a_and_b}
u\frac{\Psi'_t}{B_0}\frac{4\pi^2}{V'}
\left(
\partial_\zeta + \iota\partial_\theta
\right)
H_j
-
\nu_{zi}\frac{T}{m_z}\partial_u^2 H_j
=
\tilde S_j,
\end{eqnarray}
with $j=a,b$. Using the Fourier expansions
\begin{equation}\label{eq:Fourier_expansion_H}
H_j = \sum_{m,n =-\infty}^\infty H_{j,mn} e^{i(m\theta + n\zeta)}
\end{equation}
and
\begin{equation}
\tilde S_j = \sum_{m,n =-\infty}^\infty {\tilde S}_{j,mn} e^{i(m\theta + n\zeta)},
\end{equation}
and defining the non-dimensional parameter
\begin{equation}\label{eq:def_c}
c = \left(\frac{8\pi^2 \Psi'_t R_0 \left(
n + \iota m
\right)
}{B_0 V'}\right)^{1/3},
\end{equation}
equation \eq{eq:dke_z_plateau_3_a_and_b} translates into
\begin{eqnarray}\label{eq:dke_z_plateau_Fourier_b}
\frac{i}{v_{tz}}u c^3 H_{j,mn}
-
\nu_{zi*}
v_{tz}^2\partial_u^2 H_{j,mn}
= \frac{2 R_0}{v_{tz}}{\tilde S}_{j,mn}.
\end{eqnarray}

Making the change of variable (we do not change the name of $H_{j,mn}$)
\begin{equation}
y = \nu_{zi*}^{-1/3} c \frac{u}{v_{tz}}, 
\end{equation}
we find
\begin{eqnarray}
i y
H_{j,mn}
-
\partial_y^2 H_{j,mn}
=
\frac{2R_0}{\nu_{zi*}^{1/3}c^2 v_{tz}}\tilde S_{j,mn}.
\end{eqnarray}
The solution to this equation is
\begin{equation}\label{eq:solution_g_mn_preliminary}
H_{j,mn} = \frac{2R_0}{\nu_{zi*}^{1/3}c^2 v_{tz}}\tilde S_{j,mn}\int_0^\infty
\exp\left(-\frac{1}{3} z^3 - i y z \right)
\dd z
\end{equation}
and, employing the variable $u$,
\begin{equation}\label{eq:solution_g_mn}
H_{j,mn} = \frac{2R_0}{\nu_{zi*}^{1/3}c^2 v_{tz}}\tilde S_{j,mn}\int_0^\infty
\exp\left(-\frac{1}{3} z^3 - \frac{ic}{\nu_{zi*}^{1/3} v_{tz}} u z \right)
\dd z.
\end{equation}

\subsubsection{Explicit expression for the impurity flux in the plateau regime.}
\label{sec:explicit_expression_impurity_flux_plateau_regime}

We can obtain an explicit expression for $\Gamma_z[G_{z,p}]$. Using coordinates $u$ and $\mu$ and to lowest order in $\epsilon$,
\begin{equation}
\fl\Gamma_z[G_{z,p}] =
\frac{B_0}{2\pi}
\int_0^{2\pi}\dd\theta
\int_0^{2\pi}\dd\zeta
\int_0^\infty
\dd\mu
\int_{-\infty}^\infty \dd u \bv_d\cdot\nabla r (H_a + H_b).
\end{equation}

Now, we employ the Fourier expansions \eq{eq:Fourier_expansion_H} and
\begin{equation}\label{eq:Fourier_expansion_radial_drift}
\bv_d\cdot\nabla r = \sum_{m,n =-\infty}^\infty (\bv_d\cdot\nabla r)_{mn} e^{i(m\theta + n\zeta)}.
\end{equation}
Then,
\begin{equation}\label{eq:Gamma_z_plateau_almost_final}
\fl\Gamma_z[G_{z,p}] =
2\pi B_0
\sum_{m,n =-\infty}^\infty
\int_0^\infty
\dd\mu
\int_{-\infty}^\infty \dd u 
(\bv_d\cdot\nabla r)_{-m,-n}
(H_{a,mn} + H_{b,mn}).
\end{equation}
Here, the approximation \eq{eq:v_d_simplified} to the radial drift is understood, so that
\begin{eqnarray}
\fl
(\bv_d\cdot\nabla\rcoor)_{mn}
=
\frac{i 4\pi^2 m_z \mu}{B_0^2V' Z_z e}
\left(
I_t n B_{1,mn}
- I_p m B_{1,mn}
\right)
\nonumber\\[5pt]
\fl\hspace{1cm}
+\frac{i4\pi^2}{B_0^2V'}
(I_t n \varphi_{1,mn} - I_p m \varphi_{1,mn}),
\end{eqnarray}
with
\begin{equation}
B_1 = \sum_{m,n =-\infty}^\infty B_{1,mn} e^{i(m\theta + n\zeta)},
\end{equation}
\begin{equation}
\varphi_1 = \sum_{m,n =-\infty}^\infty \varphi_{1,mn} e^{i(m\theta + n\zeta)}.
\end{equation}

At this point, we note the identity
\begin{equation}
\lim_{k\to 0^+}\frac{1}{k}
\int_0^\infty
e^{-z^3/3}\cos\left(\frac{1}{k} x z\right)
\dd z = \pi \delta(x),
\end{equation}
which implies that the component of $H_{j,mn}$ that is even in $u$ (the odd component does not give radial transport), $H_{j,mn}^{\rm even}$, is proportional to a delta function for small collisionality,
\begin{equation}\label{eq:Heven_Dirac_delta}
\lim_{\nu_{zi*}\to 0} H_{j,mn}^{\rm even} =  \frac{2\pi R_0}{|c|^3}\tilde S_{j,mn} \delta(u).
\end{equation}
Using this and \eq{eq:def_c} in \eq{eq:Gamma_z_plateau_almost_final}, we get
\begin{equation}\label{eq:Gamma_z_plateau_final}
\fl\Gamma_z[G_{z,p}] =
\frac{B_0^2V'}{2 \Psi'_t}
\sum_{m,n =-\infty}^\infty
\frac{1}{ | n +\iota m|}
\int_0^\infty
(\bv_d\cdot\nabla r)_{-m,-n}(\tilde S_{a,mn} + \tilde S_{b,mn})
\dd\mu
,
\end{equation}
where we have assumed $\Psi'_t > 0$. Recall the decomposition \eq{eq:decomposition_Ha_Hb_plateau} and that $\hat G_{z,p}$ is $G_{z,p}$ expressed in different coordinates. Clearly, equation \eq{eq:Gamma_z_plateau_final} is the sum of a contribution from $H_a$, that we denote by $\Gamma_{z,a}\equiv\Gamma_z[H_a]$, 
\begin{equation}\label{eq:Gamma_z_plateau_final_a}
\fl\Gamma_{z,a}=
\frac{B_0^2V'}{2 \Psi'_t}
\sum_{m,n =-\infty}^\infty
\frac{1}{ | n +\iota m|}
\int_0^\infty
(\bv_d\cdot\nabla r)_{-m,-n} \tilde S_{a,mn}
\dd\mu
,
\end{equation}
and a contribution from $H_b$, that we denote by $\Gamma_{z,b}\equiv\Gamma_z[H_b]$,
\begin{equation}\label{eq:Gamma_z_plateau_final_b}
\fl\Gamma_{z,b}=
\frac{B_0^2V'}{2 \Psi'_t}
\sum_{m,n =-\infty}^\infty
\frac{1}{ | n +\iota m|}
\int_0^\infty
(\bv_d\cdot\nabla r)_{-m,-n} \tilde S_{b,mn}
\dd\mu
.
\end{equation}
We recall that $\tilde S_a$ and $\tilde S_b$ are defined in \eq{eq:source_layer_plateau_a} and \eq{eq:source_layer_plateau_b}, respectively.

The integrals over $\mu$ in \eq{eq:Gamma_z_plateau_final_a} and \eq{eq:Gamma_z_plateau_final_b} can be taken analytically. Doing this and rearranging terms, one can write $\Gamma_{z,a}$ as
\begin{eqnarray}
\Gamma_{z,a} = -\eta
\left(
D_\eta \frac{\eta'}{\eta} + D_{\varphi_0}\frac{e\varphi'_0}{T} + D_T\frac{T'}{T} + D_{n_i} \frac{n'_i}{n_i}
\right),
\end{eqnarray}
where
\begin{eqnarray}
\fl
D_\eta
=
\frac{2^{3/2}\pi^{5/2} m_z^{1/2} T^{3/2}}{Z_z^2e^2 B_0^3 V' \Psi'_t}
\sum_{m,n = -\infty}^\infty
\frac{(n I_t - m I_p)^2}{| n+\iota m |}
\Bigg[
2
\left|
\frac{B_{1,mn}}{B_0}
\right|^2
\nonumber\\[5pt]
\fl\hspace{1cm}
+
\frac{B_{1,mn}}{B_0}\frac{Z_z e \varphi^*_{1,mn}}{T}
+
\frac{B^*_{1,mn}}{B_0}\frac{Z_z e \varphi_{1,mn}}{T}
+
\left|
\frac{Z_z e \varphi_{1,mn}}{T}
\right|^2
\Bigg],
\end{eqnarray}
\begin{eqnarray}
\fl
D_{\varphi_0}
=
\frac{2^{3/2}\pi^{5/2} m_z^{1/2} T^{3/2}}{Z_z e^2 B_0^3 V'}
\sum_{m,n = -\infty}^\infty
\Bigg[
\frac{(n I_t - m I_p)^2}{\Psi'_t | n+\iota m |}
-
\frac{f_s + \left\langle B^2 u \right\rangle}{1-f_c}
\frac{n+\iota m}{| n + \iota m |}(n I_t - m I_p)
\Bigg]
\times
\nonumber\\[5pt]
\fl\hspace{1cm}
\Bigg[
2
\left|
\frac{B_{1,mn}}{B_0}
\right|^2
+
\frac{B_{1,mn}}{B_0}\frac{Z_z e \varphi^*_{1,mn}}{T}
+
\frac{B^*_{1,mn}}{B_0}\frac{Z_z e \varphi_{1,mn}}{T}
+
\left|
\frac{Z_z e \varphi_{1,mn}}{T}
\right|^2
\Bigg]
,
\end{eqnarray}
\begin{eqnarray}
\fl
D_T
=
\frac{2^{3/2}\pi^{5/2} m_z^{1/2} T^{3/2}}{Z_z e^2 B_0^3 V'}
\sum_{m,n = -\infty}^\infty
\Bigg\{
\frac{1}{2Z_z}\frac{(n I_t - m I_p)^2}{\Psi'_t | n+\iota m |}
\times
\nonumber\\[5pt]
\fl\hspace{1cm}
\Bigg[
6
\left|
\frac{B_{1,mn}}{B_0}
\right|^2
+
\frac{B_{1,mn}}{B_0}\frac{Z_z e \varphi^*_{1,mn}}{T}
+
\frac{B^*_{1,mn}}{B_0}\frac{Z_z e \varphi_{1,mn}}{T}
-
\left|
\frac{Z_z e \varphi_{1,mn}}{T}
\right|^2
\Bigg]
\nonumber\\[5pt]
\fl\hspace{1cm}
+
0.17
\frac{1}{Z_i}
\frac{f_s + \left\langle B^2 u \right\rangle}{1-f_c}
\frac{n+\iota m}{| n + \iota m |}(n I_t - m I_p)
\times
\nonumber\\[5pt]
\fl\hspace{1cm}
\Bigg[
2
\left|
\frac{B_{1,mn}}{B_0}
\right|^2
+
\frac{B_{1,mn}}{B_0}\frac{Z_z e \varphi^*_{1,mn}}{T}
+
\frac{B^*_{1,mn}}{B_0}\frac{Z_z e \varphi_{1,mn}}{T}
+
\left|
\frac{Z_z e \varphi_{1,mn}}{T}
\right|^2
\Bigg]
\Bigg\},
\end{eqnarray}
\begin{eqnarray}
\fl
D_{n_i}
=
-
\frac{2^{3/2}\pi^{5/2} m_z^{1/2} T^{3/2}}{Z_i Z_z e^2 B_0^3 V'}
\frac{f_s + \left\langle B^2 u \right\rangle}{1-f_c}
\sum_{m,n = -\infty}^\infty
\frac{n+\iota m}{| n + \iota m |}(n I_t - m I_p)
\times
\nonumber\\[5pt]
\fl\hspace{1cm}
\Bigg[
2
\left|
\frac{B_{1,mn}}{B_0}
\right|^2
+
\frac{B_{1,mn}}{B_0}\frac{Z_z e \varphi^*_{1,mn}}{T}
+
\frac{B^*_{1,mn}}{B_0}\frac{Z_z e \varphi_{1,mn}}{T}
+
\left|
\frac{Z_z e \varphi_{1,mn}}{T}
\right|^2
\Bigg]
\end{eqnarray}
and an asterisk denotes complex conjugation.

As for $\Gamma_{z,b}$, we find
\begin{eqnarray}
\fl
\Gamma_{z,b}
=
-\eta
\frac{
i \pi^{1/2} m_z^{3/2} \nu_{zi}
}
{
\sqrt{2}\, Z_z e B_0 T^{1/2} \Psi'_t
}
\sum_{m,n = -\infty}^\infty
\frac{n I_t - m I_p}{| n + \iota m |}
\times
\nonumber\\[5pt]
\fl\hspace{1cm}
\Bigg[
\left(
2 \frac{B^*_{1,mn}}{B_0} + \frac{Z_z e \varphi^*_{1,mn}}{T}
\right)
\left(
\matM:(\matI - \bun\bun) - \left\langle \matM:(\matI - \bun\bun) \right\rangle
\right)_{mn}
\nonumber\\[5pt]
\fl\hspace{1cm}
-
\left(
\frac{B^*_{1,mn}}{B_0} + \frac{Z_z e \varphi^*_{1,mn}}{T}
\right)
\left(
{\rm tr}\matM - \left\langle {\rm tr}\matM \right\rangle
\right)_{mn}
\Bigg],
\end{eqnarray}
with
\begin{eqnarray}
\fl
\matM:(\matI - \bun\bun) - \left\langle \matM:(\matI - \bun\bun) \right\rangle =
\nonumber\\[5pt]
\fl\hspace{1cm}
\sum_{m,n =-\infty}^\infty 
\left(
\matM:(\matI - \bun\bun) - \left\langle \matM:(\matI - \bun\bun) \right\rangle
\right)_{mn}
 e^{i(m\theta + n\zeta)}
\end{eqnarray}
and
\begin{eqnarray}
\fl
{\rm tr}\matM - \left\langle {\rm tr}\matM \right\rangle =
\sum_{m,n =-\infty}^\infty 
\left(
{\rm tr}\matM - \left\langle {\rm tr}\matM \right\rangle
\right)_{mn}
 e^{i(m\theta + n\zeta)}.
\end{eqnarray}

\subsubsection{Size of the impurity flux in the plateau regime.}
\label{sec:size_impurity_flux_plateau_regime}

From the expressions for $\Gamma_{z,a}$ and $\Gamma_{z,b}$ found in subsection \ref{sec:explicit_expression_impurity_flux_plateau_regime}, we learn that
\begin{equation}\label{eq:Gammaa_plateau}
\Gamma_{z,a} \sim \frac{Z_z}{\epsilon} \rho_{z*}^2 n_z v_{tz}
\end{equation}
and
\begin{equation}
\Gamma_{z,b} \sim \epsilon^{1/2} \rho_{z*} \nu_{zi*}\frac{h_{i+}}{F_{Mi}}n_z v_{tz}.
\end{equation}
In \eq{eq:Gammaa_plateau}, the factor $Z_z/\epsilon$ comes from the estimate \eq{eq:def_Upsilon} for $\Upsilon$. Note also that in \eq{eq:source_layer_plateau_a} the term containing $\Upsilon$ has the same typical size as the term proportional to $A_0$; hence, in principle, temperature screening is possible.

If the main ions are in the $1/\nu$ regime, then
\begin{equation}\label{eq:Gammab_plateau_1overnu}
\Gamma_{z,b} := \Gamma_{z}[H_b] \sim \epsilon^{1/2} Z_z^3 \frac{m_i}{m_z} \rho_{z*}^2 n_z v_{tz}.
\end{equation}

Taking the ratio of \eq{eq:Gammab_plateau_1overnu} to \eq{eq:Gammaa_plateau} we deduce that, 
\begin{equation}
\frac{\Gamma_{z,b}}{\Gamma_{z,a}}
\sim \epsilon^{3/2} Z_z^2 \frac{m_i}{m_z}.
\end{equation}
Hence, when the impurities are in the plateau regime and the main ions are in the $1/\nu$ regime, transport driven by main ion pressure anisotropy is significant if
\begin{equation}
\aleph \gtrsim 1,
\end{equation}
where $\aleph$ has been defined in \eq{eq:def_tau}.

Finally, note that (recall \eq{eq:Gammaz_Gz12})
\begin{equation}
\frac{
\Gamma_z[g_{z}^{(2)}]
}
{\Gamma_{z,b}} \sim \nu_{zi*} \ll 1,
\end{equation}
so that transport due to passing particles with large parallel velocity is negligible, as announced above.

\section{Impurities in the $1/\nu$ regime}
\label{sec:impurities_1overnu_regime}

Let us assume that
\begin{equation}
\nu_{zi*} \ll \epsilon^{3/2}.
\end{equation}
In this case, the parallel streaming term is much larger than the collision term in \eq{eq:DKE_impurities_more_explicit} for both passing and trapped particles. Therefore, one can average over the motion along magnetic field lines, and it is convenient to choose spatial coordinates in which such averaged is performed in a transparent way. We use $(r,\alpha,l)$, where $\alpha = \theta - \iota\zeta \in[0, 2\pi)$ is an angular coordinate labeling magnetic field lines and $l \in[0,l_{\rm max}(r,\alpha))$ is the arc length along the field line. Then, the magnetic field can be written as (cf. equation \eq{eq:B_contrav})
\begin{equation}
\bB = \Psi'_t \nabla r \times\nabla\alpha.
\end{equation}

Expanding $F_{z1} = F_{z1}^{(-1)} + F_{z1}^{(0)} + \dots$, with $F_{z1}^{(k)} \sim \nu_{zi*}^{k}\rho_{z*}F_{Mz}$, equation \eq{eq:DKE_impurities_more_explicit} gives, to lowest order,
\begin{equation}
\partial_l F_{z1}^{(-1)} = 0.
\end{equation}
The distribution $F_{z1}^{(-1)}$ can be taken to be zero for passing particles in large aspect ratio stellarators and/or stellarators with a sufficiently high degree of optimization (see references \cite{Calvo2017, Calvo2018, Helander2017b}). For trapped particles, $F_{z1}^{(-1)}(r,\alpha,\cE,\mu)$ is found by orbit-averaging the drift-kinetic equation to next order in the $\nu_{zi*}$ expansion,
\begin{eqnarray}\label{eq:DKEimpurities_Fminus1}
\fl
\int_{l_{b1}}^{l_{b2}} \frac{1}{|v_{||}|} 
{\cal K}F_{z1}^{(-1)}
 \dd l
 =
 \\[5pt]
 \fl\hspace{0.5cm}
 F_{Mz}
\int_{l_{b1}}^{l_{b2}} \frac{1}{|v_{||}|}
\Bigg[
\frac{1}{\nu_{zi}}
\bv_d \cdot\nabla r
\Upsilon
+
\frac{m_z}{T}
\left(\tr(\matM)  - \frac{m_z}{T} \matM \, : \left[
v_{||}^2\bun\bun + \frac{v_\perp^2}{2}\left(\matI - \bun\bun\right)
\right]
\right)
\Bigg]
 \dd l,
 \nonumber
\end{eqnarray}
where we have used that the term proportional to $A$ in the collision operator \eq{eq:C_zi_lin_expanded_in_mass_ratio} is odd in $v_{||}$ and orbit-averages to zero.

When $\varphi_1$ is neglected, it is known that terms containing derivatives with respect to the pitch angle coordinate, $\lambda = 2\mu / v^2$, dominate the differential piece of the collision operator (for us, ${\cal K}$) when $\epsilon \ll 1$~\cite{Beidler2011, Calvo2018}, and the collision term simplifies significantly. However, when $\varphi_1$ is included, $\lambda$ is not a good coordinate to determine whether a particle is trapped or passing, or to identify the largest piece of the collision operator. In our case, the operator ${\cal K}$ can still be written as a differential operator in a single variable, but this variable is not the pitch angle. We explain this in subsection \ref{sec:coll_operator_small_epsilon}.

\subsection{Collision operator for large aspect ratio}
\label{sec:coll_operator_small_epsilon}

Let us take velocity coordinates $(\hat\cE, \mu, \phi)$, where $\hat\cE$ is defined as follows. We introduce the function $U$,
\begin{equation}
U(r,\alpha,l,\mu) = \mu B(r,\alpha,l) + Z_z e(\varphi_0(r) + \varphi_1(r,\alpha,l))/m_z,
\end{equation}
and define $U_{\rm M}(r,\mu)$ as the maximum value of $U$ on the magnetic surface $r$ for a given value of $\mu$. Then, $\hat\cE$ is defined by
\begin{equation}
\hat\cE = \cE - U_{\rm M}(r,\mu).
\end{equation}
The coordinate $\mu$ takes values in $[0,\infty)$. Given a value of $\mu$, the coordinate $\hat\cE$ takes values in the interval $[U_{\rm m}(r,\mu) - U_{\rm M}(r,\mu),\infty)$, where $U_{\rm m}(r,\mu)$ is the minimum value of $U$ on the magnetic surface $r$ for that particular value of $\mu$. 

In terms of $\hat\cE$, distinguishing between passing and trapped particles is simple. If $\hat\cE > 0$, the particle is passing. The particle is trapped if
\begin{equation}\label{eq:range_hatE}
U_{\rm m}(r,\mu) - U_{\rm M}(r,\mu) \le \hat\cE \le 0.
\end{equation}

Denote by $\hat F_{z1}^{(-1)}$ the distribution $F_{z1}^{(-1)}$ expressed as a function of $(r,\alpha,\hat\cE,\mu)$. In coordinates $\hat\cE$, $\mu$ and $\phi$, the operator ${\cal K}$ reads
\begin{eqnarray}
\fl
{\cal K} \hat F_{z1}^{(-1)} = \frac{T}{m_z}\frac{1}{J}
\Bigg\{
\partial_{\hat\cE}
\left[
J F_{Mz} \nabla_v\hat\cE \cdot\nabla_v\left(\frac{\hat F_{z1}^{(-1)}}{F_{Mz}}\right)
\right]
\nonumber\\[5pt]
\fl
\hspace{1cm}
+
\partial_{\mu}
\left[
J F_{Mz} \nabla_v \mu \cdot\nabla_v\left(\frac{\hat F_{z1}^{(-1)}}{F_{Mz}}\right)
\right]
+
\partial_{\phi}
\left[
J F_{Mz} \nabla_v\phi\cdot\nabla_v\left(\frac{\hat F_{z1}^{(-1)}}{F_{Mz}}\right)
\right]
\Bigg\}.
\end{eqnarray}
Here,
\begin{equation}
\nabla_v\hat\cE = v_{||} \bun + \left(1-\frac{B_M}{B}\right)\bv_\perp,
\end{equation}
\begin{equation}
\nabla_v \mu = \frac{\bv_\perp}{B}
\end{equation}
and
\begin{equation}
\nabla_v\phi = \frac{\bun\times\bv}{v_\perp^2},
\end{equation}
with $\bv_\perp = \bv - v_{||}\bun$ and $B_M = B(r,\alpha_M, l_M)$, where $\alpha_M$ and $l_M$ correspond to the point on the magnetic surface where $U = U_M$. The Jacobian, $J$, is given by
\begin{equation}
J = \left|\nabla_v\hat\cE\cdot\left(
\nabla_v \mu\times\nabla_v\phi
\right)\right|^{-1} = \frac{B}{|v_{||}|}\, .
\end{equation}

Using that $\partial_\phi (\hat F_{z1}^{(-1)}/F_{Mz}) = 0$, $\nabla_v\hat\cE\cdot\nabla_v\phi = 0$ and $\nabla_v \mu\cdot\nabla_v\phi = 0$, we find
\begin{eqnarray}\label{eq:Browninan_motion_operator_gyrophaseinvfunc}
\fl
{\cal K} \hat F_{z1}^{(-1)} = \frac{T}{m_z}\frac{|v_{||}|}{B}
\Bigg\{
\partial_{\hat\cE}
\left[
\frac{B}{|v_{||}|} F_{Mz}
\left(
 |\nabla_v\hat\cE|^2 \partial_{\hat\cE}\left(\frac{\hat F_{z1}^{(-1)}}{F_{Mz}}\right)
 +
 \nabla_v\hat\cE \cdot \nabla_v \mu \partial_\mu\left(\frac{\hat F_{z1}^{(-1)}}{F_{Mz}}\right)
 \right)
\right]
\nonumber\\[5pt]
\fl
\hspace{1cm}
+
\partial_{\mu}
\left[
\frac{B}{|v_{||}|} F_{Mz}
\left(
 \nabla_v\hat\cE \cdot \nabla_v \mu \partial_{\hat\cE}\left(\frac{\hat F_{z1}^{(-1)}}{F_{Mz}}\right)
 +
 |\nabla_v \mu|^2 \partial_{\mu}\left(\frac{\hat F_{z1}^{(-1)}}{F_{Mz}}\right)
 \right)
\right]
\Bigg\}.
\end{eqnarray}
In this expression and the expressions that follow in this section, $v_{||}$, $F_{Mz}$, etc. are viewed as functions of the independent coordinates $\hat\cE$ and $\mu$. In these coordinates,
\begin{equation}
|v_{||}| = \sqrt{2\left(\hat\cE + U_{\rm M}(r,\mu) - U(r,\alpha,l,\mu)\right)}\, .
\end{equation}

Recalling \eq{eq:ordering_B}, \eq{eq:ordering_varphi1} and \eq{eq:range_hatE}, we get the estimate
\begin{equation}
\hat\cE \sim \epsilon v_{tz}^2
\end{equation}
for trapped particles. Hence, $\hat F_{z1}^{(-1)}$ will typically have large derivatives with respect to $\hat\cE$,
\begin{equation}
\frac{\partial_{\hat\cE} \hat F_{z1}^{(-1)}}{\hat F_{z1}^{(-1)}} \sim \frac{1}{\epsilon v_{tz}^2},
\end{equation}
whereas it will have small derivatives with respect to $\hat\mu$,
\begin{equation}
\frac{\partial_{\hat\mu} F_{z1}^{(-1)}}{F_{z1}^{(-1)}} \sim \frac{B_0}{v_{tz}^2}.
\end{equation}

Again, using \eq{eq:ordering_B} and \eq{eq:ordering_varphi1}, we find
\begin{equation}
U - U_{\rm M} \sim \epsilon v_{tz}^2.
\end{equation}
Hence, for trapped particles,
\begin{equation}
|v_{||}| = \epsilon^{1/2} v_{tz}.
\end{equation}

The term containing two derivatives with respect to $\hat\cE$ in \eq{eq:Browninan_motion_operator_gyrophaseinvfunc} is larger than the remaining ones by a factor $1/\epsilon$. Noting that, at large aspect ratio, $|\partial_{\hat\cE} F_{Mz} / F_{Mz}| \ll |\partial_{\hat\cE} \hat F_{z1}^{(-1)} / \hat F_{z1}^{(-1)}|$ and $|\nabla_v\hat\cE|^2 \simeq v_{||}^2$, we finally arrive at
\begin{eqnarray}\label{eq:Browninan_motion_operator_smallepsilon}
\fl
{\cal K} \hat F_{z1}^{(-1)} \simeq
\frac{T}{m_z}
v_{||}
\partial_{\hat\cE}
\left(
 v_{||}
\partial_{\hat\cE} \hat F_{z1}^{(-1)}
\right)
.
\end{eqnarray}

\subsection{Impurity flux for impurities in the $1/\nu$ regime}
\label{sec:Gamma_z_1overnu}

Using \eq{eq:Browninan_motion_operator_smallepsilon}, equation \eq{eq:DKEimpurities_Fminus1} becomes
\begin{eqnarray}\label{eq:DKEimpurities_1overnu_smallepsilon}
\fl
\frac{T}{m_z}
\partial_{\hat\cE}
\left[
\left(
\int_{l_{b1}}^{l_{b2}}
|v_{||}| \dd l
\right)
\partial_{\hat\cE} F_{z1}^{(-1)}
\right]
 =
 \frac{1}{\nu_{zi}}F_{Mz}
\int_{l_{b1}}^{l_{b2}} \frac{1}{|v_{||}|}
\Bigg\{
\bv_d \cdot\nabla r
\Upsilon
\nonumber\\[5pt]
\fl\hspace{1cm}
+\nu_{zi}\frac{m_z}{T}
\left(\tr(\matM)  - \frac{m_z}{T} {\matM} :
\left[
v_{||}^2\bun\bun + \frac{v_\perp^2}{2}\left(\matI - \bun\bun\right)
\right]
\right)
\Bigg\}
 \dd l.
\end{eqnarray}

In analogy to what we did in previous sections, we write $F_{z1}^{(-1)} = \Ha + \Hb$, where $\Ha$ and $\Hb$ are the solutions of
\begin{eqnarray}\label{eq:DKEimpurities_1overnu_smallepsilon_Ha}
\fl
\frac{T}{m_z}
\partial_{\hat\cE}
\left[
\left(
\int_{l_{b1}}^{l_{b2}}
|v_{||}| \dd l
\right)
\partial_{\hat\cE}\Ha
\right]
 =
 \frac{1}{\nu_{zi}}F_{Mz}
\int_{l_{b1}}^{l_{b2}} \frac{1}{|v_{||}|}
\bv_d \cdot\nabla r
\Upsilon
 \dd l
\end{eqnarray}
and
\begin{eqnarray}\label{eq:DKEimpurities_1overnu_smallepsilon_Hb}
\fl
\frac{T}{m_z}
\partial_{\hat\cE}
\left[
\left(
\int_{l_{b1}}^{l_{b2}}
|v_{||}| \dd l
\right)
\partial_{\hat\cE}\Hb
\right]
 =
 \nonumber\\[5pt]
\fl\hspace{1cm}
F_{Mz}
\int_{l_{b1}}^{l_{b2}} \frac{1}{|v_{||}|}
\Bigg\{
\frac{m_z}{T}
\left(\tr(\matM)  - \frac{m_z}{T} {\matM} :
\left[
v_{||}^2\bun\bun + \frac{v_\perp^2}{2}\left(\matI - \bun\bun\right)
\right]
\right)
\Bigg\}
 \dd l.
\end{eqnarray}
These equations can be integrated numerically in a straightforward way.

Noting that
\begin{equation}
\bv_d \cdot\nabla r \sim \rho_{z*} v_{tz},
\end{equation}
from \eq{eq:DKEimpurities_1overnu_smallepsilon_Ha} we deduce
\begin{equation}\label{eq:size_Ha_1overnu}
\Ha \sim \frac{Z_z \rho_{z*}}{\nu_{zi*}} F_{Mz},
\end{equation}
where the factor $Z_z$ comes from the estimate \eq{eq:def_Upsilon} for $\Upsilon$.

Recalling \eq{eq:sizeM}, from \eq{eq:DKEimpurities_1overnu_smallepsilon_Hb} we get
\begin{equation}\label{eq:size_Hb_1overnu}
\Hb \sim \epsilon^{3/2}\frac{h_{i+}}{F_{Mi}} F_{Mz}.
\end{equation}

Employing \eq{eq:size_Ha_1overnu} and \eq{eq:size_Hb_1overnu} in \eq{eq:def_Gamma_z}, we reach the estimates
\begin{equation}\label{eq:Gammaa_1overnu}
\Gamma_{z,a} := \Gamma_z[\Ha] \sim \epsilon^{1/2} \frac{Z_z \rho_{z*}^2}{\nu_{zi*}} n_z v_{tz}
\end{equation}
and 
\begin{equation}
\Gamma_{z,b} := \Gamma_z[\Hb] \sim \epsilon^2 \rho_{z*} \frac{h_{i+}}{F_{Mi}} n_z v_{tz}.
\end{equation}
If the main ions are in the $1/\nu$ regime (recall \eq{eq:hi_over_FMi_1overnu}, \eq{eq:relation_rhoz_rhoi} and \eq{eq:relation_nuzi_nuii}), one gets
\begin{equation}\label{eq:Gammab_1overnu}
\Gamma_{z,b} := \Gamma_z[\Hb] \sim \epsilon^2 Z_z^3 \frac{m_i}{m_z}
\frac{\rho_{z*}^2}{\nu_{zi*}} n_z v_{tz}.
\end{equation}

The ratio of \eq{eq:Gammab_1overnu} to \eq{eq:Gammaa_1overnu} gives
\begin{equation}
\frac{\Gamma_{z,b}}{\Gamma_{z,a}} \sim \epsilon^{3/2}
Z_z^2 \frac{m_i}{m_z}.
\end{equation}
Therefore, when both the impurities and the main ions are in the $1/\nu$ regime, main ion pressure anisotropy must be taken into account if
\begin{equation}
\aleph
\gtrsim 1.
\end{equation}

Note that $\aleph \gsim 1$ requires to have highly charged impurities. Since highly charged impurities tend to be collisional, we expect our calculation for the $1/\nu$ regime to be less relevant than the calculations for the plateau and Pfirsch-Schl{\"u}ter regimes.

\section{Numerical evaluation of the analytical results}
\label{sec:numerical_evaluation}

\begin{figure}[H]
\centering
\includegraphics[width=0.9\textwidth]{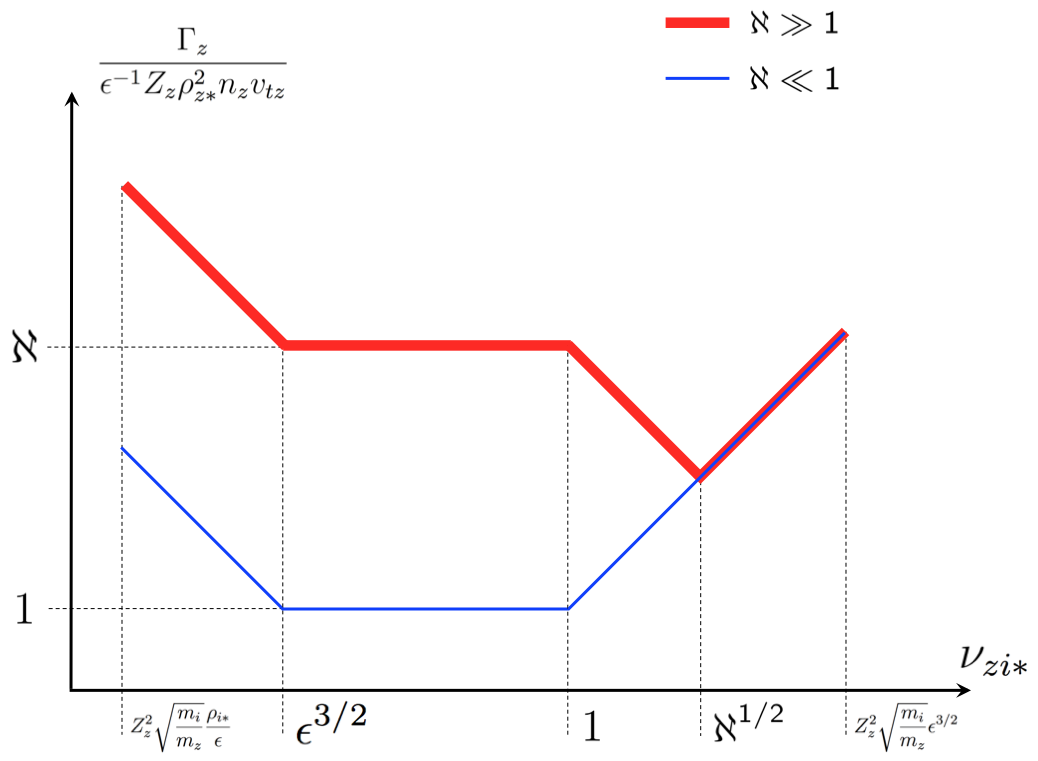}
\caption{Cartoon summarizing the results of the article for the behavior of $\Gamma_z$ versus $\nu_{zi*}$ in a log-log scale. Here, $\aleph = \epsilon^{3/2} Z_z^2 m_i / m_z$. The precise definition of the rest of the quantities can be found in the main text. The thin line, $\aleph\ll 1$, corresponds to cases in which main ion pressure anisotropy gives a negligible contribution to $\Gamma_z$. The thick line corresponds to $\aleph\gg 1$. For $1\ll \nu_{zi*} \ll \aleph^{1/2}$, the thick line reflects the $\nu_{zi*}^{-1}$ scaling of $\Gamma_{z,b}$ (see subsection \ref{sec:comparison_Gammaa_vs_Gammab_PS}). For $\nu_{zi*}\gg \aleph^{1/2}$, $\Gamma_{z,a}$ always dominates and $\Gamma_z$ scales with $\nu_{zi*}$. In this cartoon, main ions are assumed to be always in the $1/\nu$ regime. Below $\nu_{zi*} \sim Z_z^2\sqrt{m_i/m_z}\, \rho_{i*}/\epsilon$, the main ions leave the $1/\nu$ regime and enter the $\sqrt{\nu}$ regime. Above $\nu_{zi*} \sim Z_z^2\sqrt{m_i/m_z} \epsilon^{3/2}$, the main ions leave the $1/\nu$ regime and enter the plateau regime. For the small $\mathbf{E}\times\bB$ drift assumption to hold (recall the discussion around \eq{eq:Machnumber}), $\rho_{i*}$ has to satisfy $\rho_{i*} \ll \epsilon^2\sqrt{m_i/m_z}$.}
\label{fig:regimes_Gammaz}
\end{figure}

Figure \ref{fig:regimes_Gammaz} synthesizes the scalings and typical sizes for $\Gamma_z$ derived in previous sections for several collisionality regimes. We have found that there are two essentially different situations as far as the relevance of main ion impurity anisotropy is concerned, distinguished by the value of $\aleph = \epsilon^{3/2} Z_z^2 m_i / m_z$. If $\aleph \ll 1$, main ion pressure anisotropy is negligible. If $\aleph \gg 1$, main ion pressure anisotropy becomes the main drive for impurity transport. As remarked at the end of Section \ref{sec:impurities_1overnu_regime}, the condition $\aleph \gtrsim 1$ typically requires impurities with high electric charge, and such impurities are more likely to be in the plateau or Pfirsch-Schl{\"u}ter regimes than in the $1/\nu$ regime.

Next, we give realistic examples to illustrate the relevance of main-ion-pressure-anisotropy-driven neoclassical impurity transport by numerically evaluating the analytical expressions derived for $\Gamma_{z,a}$ and $\Gamma_{z,b}$ in the plateau and Pfirsch-Schl{\"u}ter regimes. The expression for $\Gamma_{z,a}$ in the Pfirsch-Schl{\"u}ter regime can be found in equation (25) of \cite{Calvo2018b}. The expressions for $\Gamma_{z,b}$ in the Pfirsch-Schl{\"u}ter regime and for $\Gamma_{z,a}$ and $\Gamma_{z,b}$ in the plateau regime have been derived in the present paper (see \eq{eq:Gammab_PS_explicit_result} and \eq{eq:Gamma_z_plateau_final}).

As trace impurity, we take the charge state $Z_z = 40$ of tungsten and we set $\eta'=0$. The main ion species is deuterium. We will perform a scan in the density of the main ions, while keeping constant the rest of the simulation parameters, in particular $n_i'/n_i$. For each point in the scan, we solve the drift-kinetic equation of the main ions using the code \texttt{KNOSOS}~\cite{Velasco2019}. In these examples, we set $\varphi_1=0$. We emphasize that \texttt{KNOSOS} solves bounce-averaged equations and can calculate $h_{i+}$, but not $h_{i-}$. This code is employed here to evaluate $\Gamma_{z,b}$, which is the component of the impurity flux that depends on $h_{i+}$.

\begin{figure}[H]
\centering
\includegraphics[angle=0,width=0.8\columnwidth]{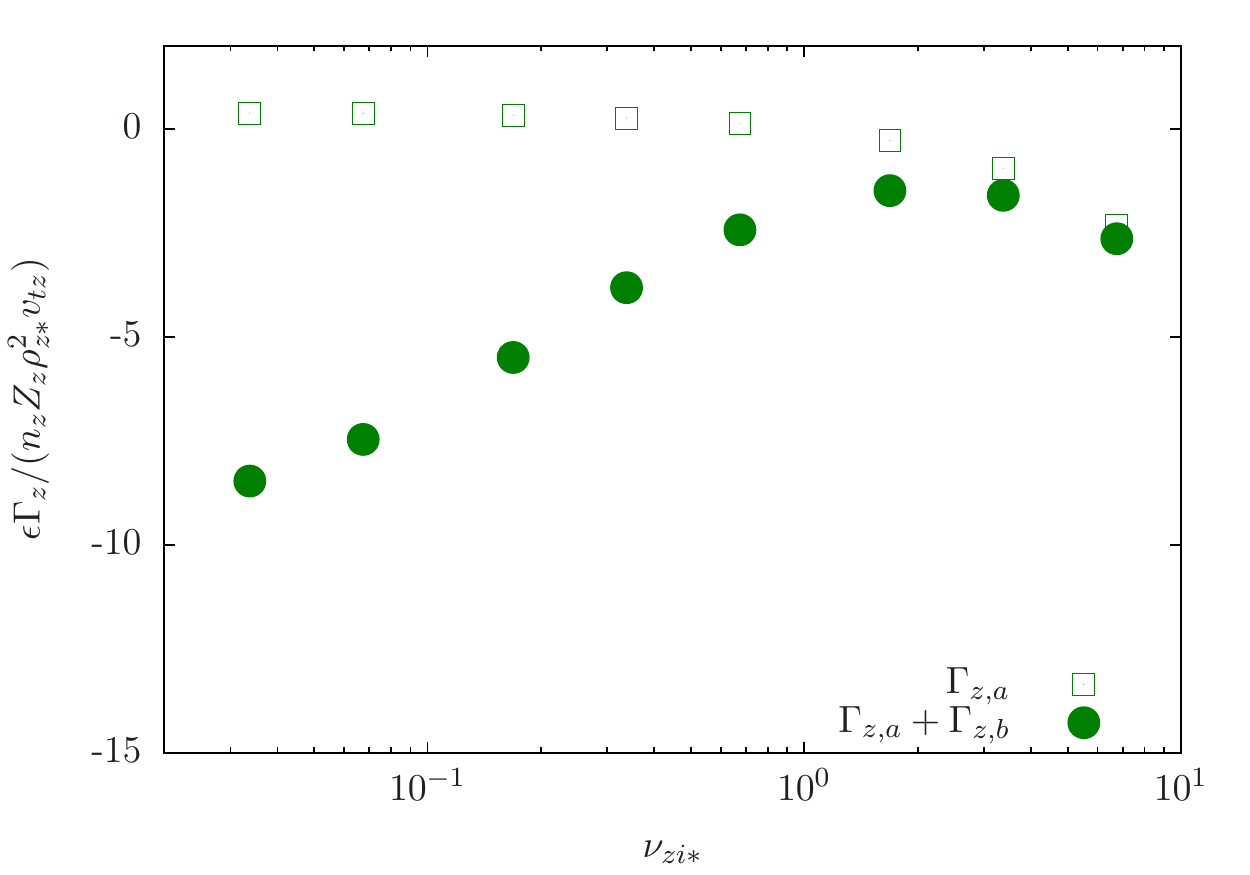}
\includegraphics[angle=0,width=0.8\columnwidth]{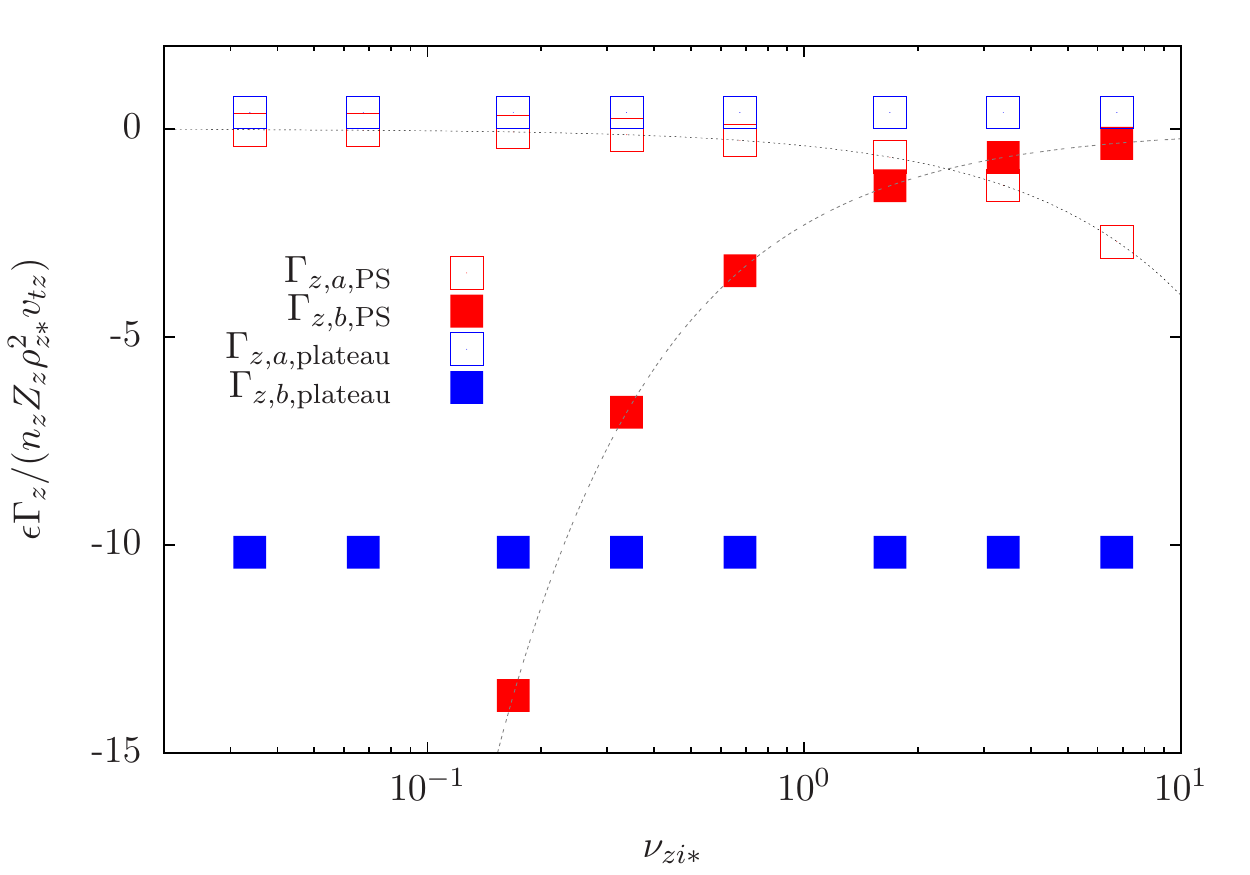}
\caption{Numerical evaluation of the impurity flux for the LHD plasma described in the text. Note that a linear scale in the vertical axis and a logarithmic scale in the horizontal axis are employed. Figure \ref{fig:example_LHD}(top) shows a comparison of the flux including (full circles) and excluding (empty squares) the effect of main ion pressure anisotropy. Figure \ref{fig:example_LHD}(bottom) shows the values of $\Gamma_{z,a}$ and $\Gamma_{z,b}$ in the asymptotic plateau and Pfirsch-Schl{\"u}ter regimes employed to produce figure \ref{fig:example_LHD}(top). The total flux $\Gamma_{z,a}$ is obtained as the sum of the plateau and Pfirsch-Schl\"uter contributions to $\Gamma_{z,a}$. As for $\Gamma_{z,b}$, we take $\Gamma_{z,b}^{-1} = \Gamma_{z,b, {\rm plateau}}^{-1} + \Gamma_{z,b, {\rm PS}}^{-1}$, where $\Gamma_{z,b, {\rm plateau}}$ and $\Gamma_{z,b, {\rm PS}}$ are the plateau and Pfirsch-Schl\"uter contributions to $\Gamma_{z,b}$. For reference, the dashed line represents a linear dependence on $\nu_{zi*}$ and the solid curve is proportional to $\nu_{zi*}^{-1}$.}
\label{fig:example_LHD}
\end{figure}

First, we use an inward-shifted configuration of the Large Helical Device (LHD), characterized by having major radius $R_0=3.67\,$m, and we focus on the flux-surface $r/a=0.8$, where $a=0.64\,$m. The magnetic field and profiles are those of the plasma simulated in~\cite{Calvo2018b}; in particular, $T=1.3\,$keV, $aT'/T=-3.4$ and $an_i'/n_i=-2.6$. The main ion density is scanned so that the collisionality goes from $\nu_{zi*} \sim 10^{-2}$ to $\nu_{zi*} \sim 10$ and we take $\varphi'_0 = 0$. When solving the main ion drift-kinetic equation with \texttt{KNOSOS}, we have switched off the component of the magnetic drift that is tangent to the flux surface, whereas the tangential component of the $\mathbf{E} \times \mathbf{B}$ drift vanishes because $\varphi'_0 = 0$. An important consequence of not having tangential drifts is that the main ions are in the $1/\nu$ regime for all values of the collisionality in the range selected here (we come back to this below). In figure \ref{fig:example_LHD}(top), the total flux $\Gamma_{z,a} + \Gamma_{z,b}$ is compared to the flux obtained by neglecting the effect of the main ion pressure anisotropy, showing that neglecting $\Gamma_{z,b}$ would lead to an incorrect result for the impurity flux in most of the collisionality range, and that, actually, $\Gamma_z$ is dominated by $\Gamma_{z,b}$ below $\nu_{zi*} \sim 1$. Figure \ref{fig:example_LHD}(bottom) exhibits the collisionality dependence of $\Gamma_{z,a}$ and $\Gamma_{z,b}$ in the plateau and Pfirsch-Schl{\"u}ter regimes discussed in this paper. We note that, in this example, $\aleph = 1.27$.

\begin{figure}[H]
\centering
\includegraphics[angle=0,width=0.8\columnwidth]{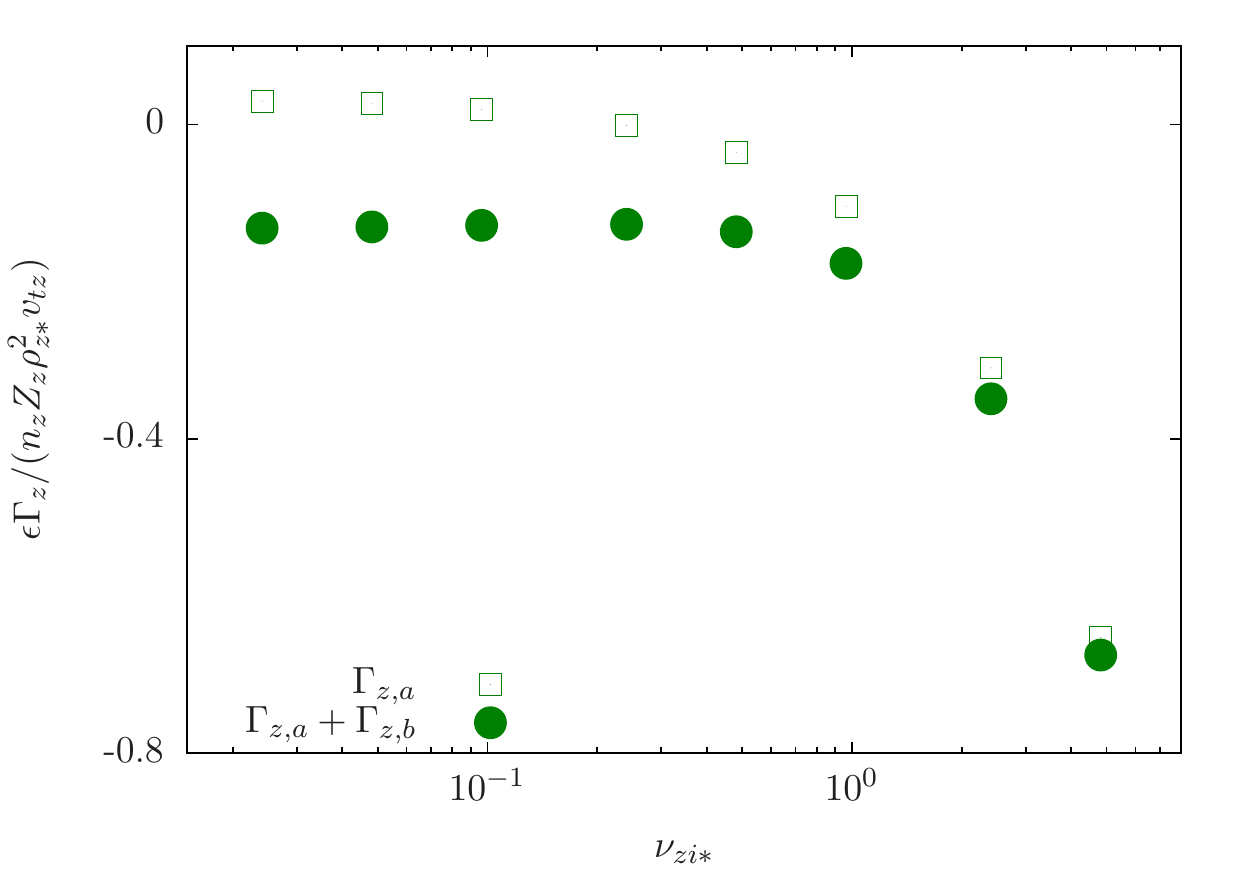}
\includegraphics[angle=0,width=0.8\columnwidth]{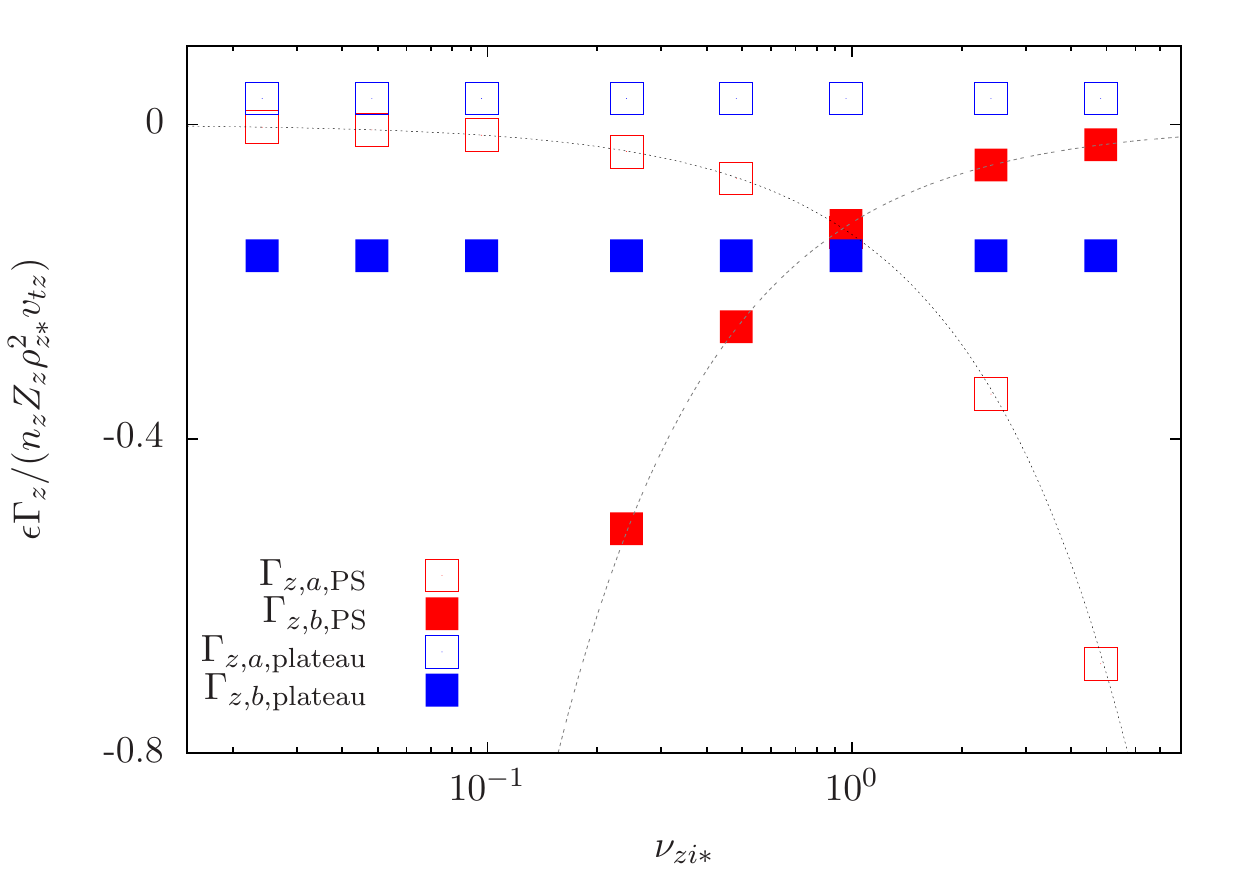}
\caption{Numerical evaluation of the impurity flux for the W7-X plasma described in the text. Note that a linear scale in the vertical axis and a logarithmic scale in the horizontal axis are employed. Here, the calculation has been carried out removing the term that involves the component of the magnetic drift tangent to the flux surface in the main ion drift-kinetic equation. Figure \ref{fig:example_W7X_without_tangential_drift}(top) shows a comparison of the flux including (full circles) and excluding (empty squares) the effect of main ion pressure anisotropy. Figure \ref{fig:example_W7X_without_tangential_drift}(bottom) shows the values of $\Gamma_{z,a}$ and $\Gamma_{z,b}$ in the asymptotic plateau and Pfirsch-Schl{\"u}ter regimes employed to produce figure \ref{fig:example_W7X_without_tangential_drift}(top). For reference, the dashed line represents a linear dependence on $\nu_{zi*}$ and the solid curve is proportional to $\nu_{zi*}^{-1}$.}
\label{fig:example_W7X_without_tangential_drift}
\end{figure}

In figure \ref{fig:example_W7X_without_tangential_drift} we give the results of an analogous calculation on the flux surface $r/a = 0.8$ of the standard configuration of W7-X, with $R_0 = 5.51\,$m and $a = 0.51\,$m. The trace impurity and main plasma parameters and profiles (normalized to the minor radius) are the same as in the LHD case. Although the results are qualitatively similar, we see that the weight of $\Gamma_{z,b}$ is smaller in W7-X due to its larger aspect ratio compared to LHD. In this example, $\aleph = 0.49$.

\begin{figure}[H]
\centering
\includegraphics[angle=0,width=0.8\columnwidth]{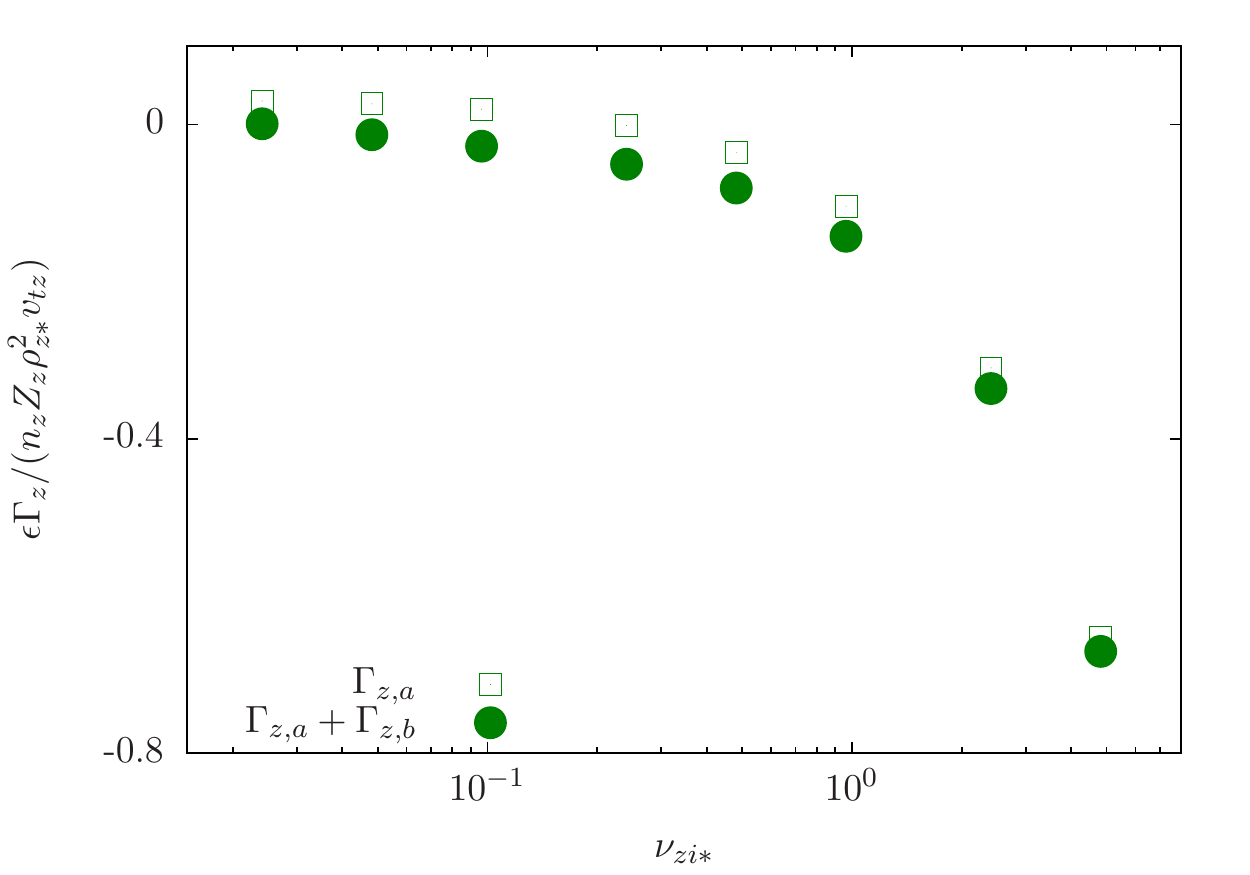}
\includegraphics[angle=0,width=0.8\columnwidth]{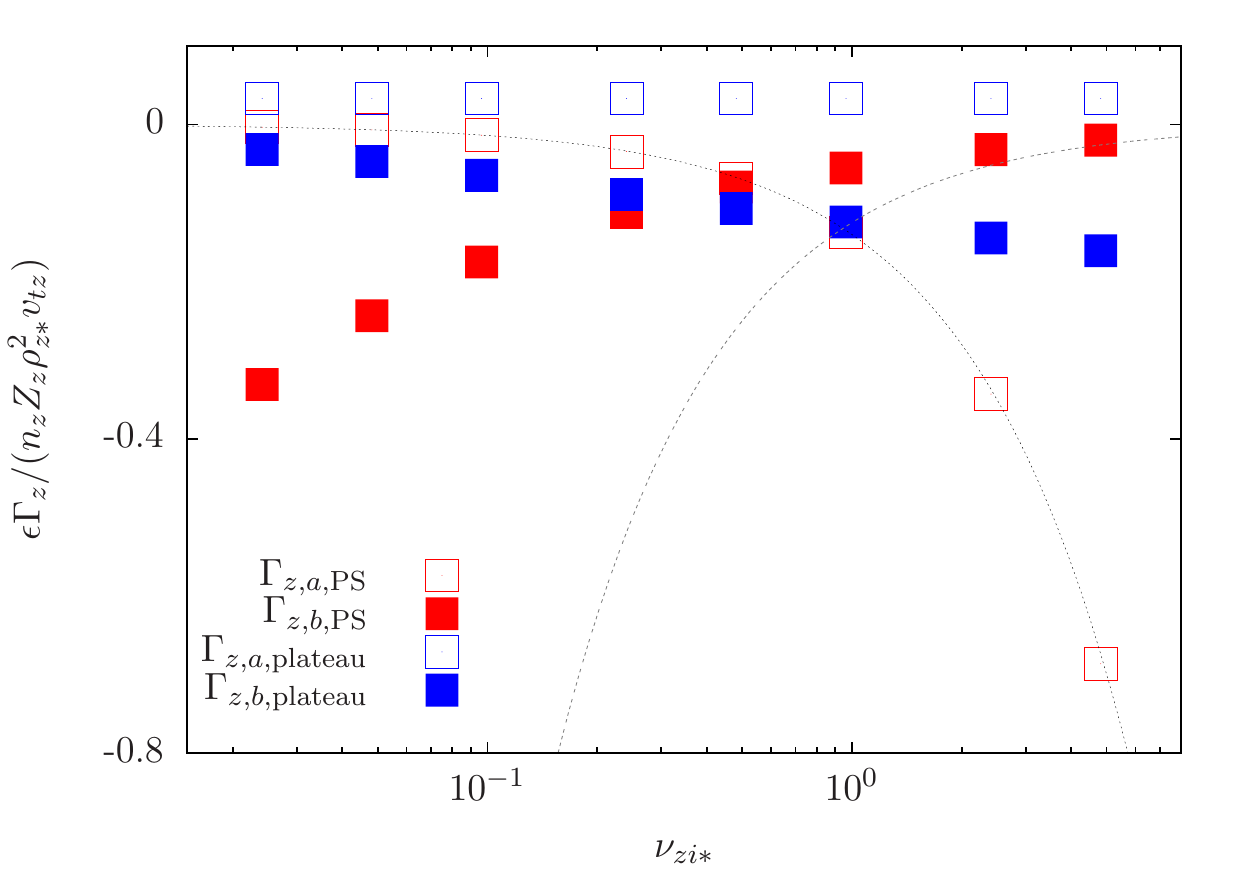}
\caption{Numerical evaluation of the impurity flux for the W7-X plasma described in the text. Note that a linear scale in the vertical axis and a logarithmic scale in the horizontal axis are employed. Here, the calculation has been carried out incorporating the term that involves the component of the magnetic drift tangent to the flux surface in the main ion drift-kinetic equation. Figure \ref{fig:example_W7X_with_tangential_drift}(top) shows a comparison of the flux including (full circles) and excluding (empty squares) the effect of main ion pressure anisotropy. Figure \ref{fig:example_W7X_with_tangential_drift}(bottom) shows the values of $\Gamma_{z,a}$ and $\Gamma_{z,b}$ in the asymptotic plateau and Pfirsch-Schl{\"u}ter regimes employed to produce figure \ref{fig:example_W7X_with_tangential_drift}(top). For reference, the dashed line represents a linear dependence on $\nu_{zi*}$ and the solid curve is proportional to $\nu_{zi*}^{-1}$.}
\label{fig:example_W7X_with_tangential_drift}
\end{figure}

In figure \ref{fig:example_W7X_with_tangential_drift} we show the result of redoing the W7-X calculations after switching on the component of the magnetic drift tangent to the flux surface in the main ion drift-kinetic equation (the effect of including the tangential drift in the impurity drift-kinetic equation is not studied in this paper). In these simulations, the main ions are not in the $1/\nu$ regime for the whole range of collisionality values. This is evident from  figure \ref{fig:example_W7X_with_tangential_drift}(bottom), where, for sufficiently low values of the collisionality, $\Gamma_{z,b}$ in the `plateau' regime is not independent of $\nu_{zi*}$ and, in the Pfirsch-Schl{\"u}ter regime, $\Gamma_{z,b}$ does not scale with the inverse of $\nu_{zi*}$. The reason is that, for collisionality low enough for the tangential drifts to be relevant, the distribution function of the main ions ceases to scale with the inverse of the collisionality~\cite{Calvo2017, Calvo2018} (the effect would be similar for an $\mathbf{E} \times \mathbf{B}$ drift caused by a relatively small radial electric field). Therefore, the relative weight of $\Gamma_{z,b}$ with respect to $\Gamma_{z,a}$ decreases compared to the case without tangential drifts (see figure \ref{fig:example_W7X_with_tangential_drift}(top) and figure \ref{fig:example_W7X_without_tangential_drift}(top)). From figure \ref{fig:example_W7X_with_tangential_drift}(top) it is clear that, for $\nu_{zi*} \lesssim 1$, neglecting $\Gamma_{z,b}$ would still be incorrect.

\section{Conclusions}
\label{sec:conclusions}

Main ion dynamics affects impurity transport via two basic mechanisms: determining the electric field and through impurity-ion collisions. In this paper, we focus on the second one.

To lowest order in a mass ratio expansion $\sqrt{m_i/m_z} \ll 1$, where $m_i$ and $m_z$ are the main ion and the impurity masses, the impurity-ion collision operator only depends on the component of the main ion distribution that is odd in the parallel velocity, $h_{i-}$. In fluid terms, using this lowest-order collision operator implies that the radial impurity flux, $\Gamma_z$, is driven only by the parallel friction of the impurities with the main ions.  However, this description might not be complete for stellarators. Whereas terms in the impurity-ion collision operator proportional to the component of the main ion distribution that is even in the parallel velocity, $h_{i+}$, are small in $\sqrt{m_i/m_z}$, they can end up driving a non-negligible amount of impurity transport because $h_{i+}$ can be large in regimes of main ion low collisionality. Hence, in principle, one can distinguish two collisional drives for stellarator impurity transport with different physical origin: parallel friction and main ion pressure anisotropy. In this article, we have derived the impurity-ion collision operator keeping terms of sufficiently high order in $\sqrt{m_i/m_z} \ll 1$ to retain the effect of main ion pressure anisotropy. With this collision operator, and using the trace impurity approximation, we have solved the impurity drift-kinetic equation in the Pfirsch-Schl{\"u}ter, plateau and $1/\nu$ regimes, assuming that the main ions have low collisionality and that the $\mathbf{E}\times\bB$ drift term in the impurity drift-kinetic equation is small compared to the parallel streaming term. For each regime, we have deduced the conditions (plasma, impurity and geometric parameters) that determine whether or not main ion pressure anisotropy gives a significant contribution to $\Gamma_z$. Finally, we have illustrated the potential importance of this effect by numerically evaluating the analytical expressions of $\Gamma_z$ for realistic stellarator plasmas.

The expressions for $\Gamma_z$ derived in this article, together with that in \cite{Calvo2018b}, and their coupling to the code \texttt{KNOSOS} (in the way described in Section \ref{sec:numerical_evaluation}) provide a numerical tool for a fast evaluation of neoclassical impurity transport in stellarators, with potential applications to the design of operation scenarios and stellarator optimization.

\ack

I.~C. and J.~L.~V. thank H.~Sugama and S.~Satake for interesting and helpful discussions. This work has been carried out within the framework of the EUROfusion Consortium and has received funding from the Euratom research and training programme 2014-2018 and 2019-2020 under grant agreement No 633053. The views and opinions expressed herein do not necessarily reflect those of the European Commission. This research was supported in part by grant ENE2015-70142-P, Ministerio de Econom\'ia y Competitividad, Spain and grant PGC2018-095307-B-I00, Ministerio de Ciencia, Innovaci\'on y Universidades, Spain.

\appendix

\section{Mass ratio expansion of the linealized impurity-ion collision operator}
\label{sec:C_zi}

Let us derive expression \eq{eq:C_zi_lin_expanded_in_mass_ratio}. The Landau operator for collisions between species $s$ and $s'$ (see, for example, \cite{Helander_book_2002}) is
\begin{eqnarray}
\fl
C_{ss'}[F_s,F_{s'}] = \gamma_{ss'}\nabla_v\cdot\left[
\int
\nabla_g\nabla_g g\cdot\left(
F_{s'}(\bv')\nabla_v F_s(\bv) - \frac{m_s}{m_{s'}} F_{s}(\bv)\nabla_{v'} F_{s'}(\bv')
\right)\dd^3v'
\right],
\end{eqnarray}
where $\bg =\bv - \bv'$,
\begin{equation}
\gamma_{ss'} = \frac{2\pi Z_s^2 Z_{s'}^2 e^4\ln\Lambda}{(4\pi\epsilon_0)^2 m_s^2}
\end{equation}
and $\ln \Lambda$ is the Coulomb logarithm.

Take two Maxwellians with the same temperature,
\begin{eqnarray}\label{eq:two_Maxwellians}
F_{Ms}(\rcoor,v) = 
n_s(\rcoor)\left(\frac{m_s}{2\pi T(\rcoor)}\right)^{3/2}
\exp\left(-\frac{m_s v^2}{2T(\rcoor)}\right),
\nonumber\\[5pt]
F_{Ms'}(\rcoor,v) = 
n_{s'}(\rcoor)\left(\frac{m_{s'}}{2\pi T(\rcoor)}\right)^{3/2}
\exp\left(-\frac{m_{s'} v^2}{2T(\rcoor)}\right),
\end{eqnarray}
and write $F_s = F_{Ms} + h_s$ and $F_{s'} = F_{Ms'} + h_{s'}$. Recalling that $C_{ss'}[F_{Ms},F_{Ms'}] = 0$ and dropping terms quadratic in the functions $h_{s}$ and $h_{s'}$, we obtain the linearized collision operator
\begin{eqnarray}\label{eq:C_ss'_lin}
\fl
C_{ss'}^{(\ell)}[h_{s};h_{s'}] = C_{ss'}[h_{s},F_{Ms'}] + C_{ss'}[F_{Ms},h_{s'}] = 
\nonumber\\[5pt]
\fl\hspace{1cm}
\gamma_{ss'}\nabla_v\cdot\left[
\int
\nabla_g\nabla_g g\cdot\left(
F_{Ms'}(\bv')\nabla_v h_{s}(\bv) - \frac{m_s}{m_{s'}} h_{s}(\bv)\nabla_{v'} F_{Ms'}(\bv')
\right)\dd^3v'
\right]
\nonumber\\[5pt]
\fl\hspace{1cm}
+
\gamma_{ss'}\nabla_v\cdot\left[
\int
\nabla_g\nabla_g g\cdot\left(
h_{s'}(\bv')\nabla_v F_{Ms}(\bv) - \frac{m_s}{m_{s'}} F_{Ms}(\bv)\nabla_{v'} h_{s'}(\bv')
\right)\dd^3v'
\right]
.
\end{eqnarray}

From here on, we select $s=z$ and $s'=i$. Hence,
\begin{eqnarray}\label{eq:C_zi_lin}
\fl
C_{zi}^{(\ell)}[h_{z};h_{i}] = C_{zi}[h_{z},F_{Mi}] + C_{zi}[F_{Mz},h_{i}] = 
\nonumber\\[5pt]
\fl\hspace{1cm}
\gamma_{zi}\nabla_v\cdot\left[
\int
\nabla_g\nabla_g g\cdot\left(
F_{Mi}(\bv')\nabla_v h_{z}(\bv) - \frac{m_z}{m_{i}} h_{z}(\bv)\nabla_{v'} F_{Mi}(\bv')
\right)\dd^3v'
\right]
\nonumber\\[5pt]
\fl\hspace{1cm}
+
\gamma_{zi}\nabla_v\cdot\left[
\int
\nabla_g\nabla_g g\cdot\left(
h_{i}(\bv')\nabla_v F_{Mz}(\bv) - \frac{m_z}{m_{i}} F_{Mz}(\bv)\nabla_{v'} h_{i}(\bv')
\right)\dd^3v'
\right]
.
\end{eqnarray}

Let us assume $\sqrt{m_i/m_z}\ll 1$. First, we focus on $C_{zi}[h_z,F_{Mi}]$. Expanding in mass ratio,
\begin{equation}\label{eq:expansion_g}
\nabla_g\nabla_g g \simeq
\nabla_{v'}\nabla_{v'}v' - \bv\cdot \nabla_{v'}\nabla_{v'}\nabla_{v'}v',
\end{equation}
so that
\begin{eqnarray}\label{eq:C_zi_lin_differential}
\fl
C_{zi}[h_z,F_{Mi}] \simeq
\gamma_{zi}\nabla_v\cdot
\Bigg[
\left(\int
\nabla_{v'}\nabla_{v'}v'
F_{Mi}(\bv')
\dd^3v'\right)\cdot \nabla_v h_{z}(\bv)
\nonumber\\[5pt]
\fl\hspace{1cm}
-
\frac{m_z}{T}
\left(
\int
\bv'\cdot\nabla_{v'}\nabla_{v'}\nabla_{v'}v' F_{Mi}(\bv')
 \dd^3v'
 \right)\cdot\bv \,
 h_{z}(\bv)
\Bigg].
\end{eqnarray}
In the second term on the right-hand side of this expression, the leading term (i.e. the first term) on the right-hand side of \eq{eq:expansion_g} has given a vanishing contribution due to  $\nabla_{v'} F_{Ms}(\bv') = -(m_s\bv'/T) F_{Ms}(\bv')$ and $\nabla_{v'}\nabla_{v'}v'\cdot\bv' = 0$. It is easy to check that both terms on the right-hand side of \eq{eq:C_zi_lin_differential} have the same size in the mass ratio expansion.

The integrals in \eq{eq:C_zi_lin_differential} can be computed analytically. Noting that
\begin{equation}
\nabla_{v'}\nabla_{v'}v' = \frac{1}{v'}\left(
\matI - \frac{\bv'\bv'}{(v')^2}
\right)
\end{equation}
and
\begin{equation}
\bv'\cdot\nabla_{v'}\nabla_{v'}\nabla_{v'}v' = -\nabla_{v'}\nabla_{v'}v',
\end{equation}
a straightforward calculation gives
\begin{equation}
\int
\nabla_{v'}\nabla_{v'}v'
F_{Mi}(\bv')
\dd^3v' = \frac{2\sqrt{2}\, n_i}{3\sqrt{\pi}}\sqrt{\frac{m_i}{T}}\matI
\end{equation}
and
\begin{equation}
\int
\bv'\cdot\nabla_{v'}\nabla_{v'}\nabla_{v'}v' F_{Mi}(\bv')
 \dd^3v' = - \frac{2\sqrt{2}\, n_i}{3\sqrt{\pi}}\sqrt{\frac{m_i}{T}}\matI.
\end{equation}
Hence, to lowest order in the $\sqrt{m_i/m_z}$ expansion,
\begin{eqnarray}\label{eq:C_zi_lin_differential_2}
\fl
C_{zi}[h_z,F_{Mi}] \simeq
\gamma_{zi}\frac{2\sqrt{2}\, n_i}{3\sqrt{\pi}}\sqrt{\frac{m_i}{T}}
\,
\nabla_v\cdot
\left(
F_{Mz}
\nabla_v
\left(
 \frac{h_{z}(\bv)}{F_{Mz}}
\right)
\right).
\end{eqnarray}

We turn to $C_{zi}[F_{Mz},h_{i}]$. Employing \eq{eq:expansion_g},
\begin{eqnarray}\label{eq:C_zi_lin_integral_aux_preliminary}
\fl
C_{zi}[F_{Mz},h_{i}] \simeq
-\gamma_{zi}\nabla_v\cdot
\Bigg[
\frac{m_z}{T}
\left(
\int
\nabla_{v'}\nabla_{v'}v' 
h_{i}(\bv')
\dd^3v'
\right)
\cdot
\bv
F_{Mz}(\bv)
\Bigg]
\nonumber\\[5pt] 
\fl\hspace{1cm}
-
\gamma_{zi}\nabla_v\cdot
\Bigg[
\frac{m_z}{m_{i}}
F_{Mz}(\bv)
\int
(
\nabla_{v'}\nabla_{v'}v' - \bv\cdot \nabla_{v'}\nabla_{v'}\nabla_{v'}v'
)\cdot
\nabla_{v'} h_{i}(\bv')
\dd^3v'
\Bigg].
\end{eqnarray}
In the first term on the right-hand side of \eq{eq:C_zi_lin_integral_aux_preliminary}, it has been enough to keep the contribution of the first term on the right-hand side of \eq{eq:expansion_g} because the second term on the right-hand side of \eq{eq:expansion_g} always gives a negligible contribution. However, in the second term on the right-hand side of \eq{eq:C_zi_lin_integral_aux_preliminary}, we have kept the contributions coming from the two terms on the right-hand side of \eq{eq:expansion_g}.

From \eq{eq:C_zi_lin_integral_aux_preliminary}, we easily obtain
\begin{eqnarray}\label{eq:C_zi_lin_integral_aux}
\fl
C_{zi}[F_{Mz},h_{i}] \simeq
-\gamma_{zi}\frac{m_z}{T}
\left(\int
\nabla_{v'}^2 v' 
h_{i}(\bv')
\dd^3v'
\right)
F_{Mz}(\bv)
\nonumber\\[5pt] 
\fl\hspace{1cm}
+\gamma_{zi}
\frac{m_z^2}{T^2}\left(\int
\nabla_{v'}\nabla_{v'}v' 
h_{i}(\bv')
\dd^3v'
\right):
\bv\bv
F_{Mz}(\bv)
\nonumber\\[5pt] 
\fl\hspace{1cm}
+\gamma_{zi}
\frac{m_z^2}{m_i T}\left(\int
\nabla_{v'}\nabla_{v'}v'\cdot\nabla_{v'}
h_{i}(\bv')
\dd^3v'
\right)\cdot
\bv
F_{Mz}(\bv)
\nonumber\\[5pt] 
\fl\hspace{1cm}
+\gamma_{zi}
\frac{m_z}{m_i }\left(\int
\nabla_{v'}\nabla_{v'}^2 v'\cdot\nabla_{v'}
h_{i}(\bv')
\dd^3v'
\right)
F_{Mz}(\bv)
\nonumber\\[5pt] 
\fl\hspace{1cm}
-\gamma_{zi}
\frac{m_z^2}{m_i T}\left(\int
\nabla_{v'}\nabla_{v'}\nabla_{v'} v'\cdot\nabla_{v'}
h_{i}(\bv')
\dd^3v'
\right):\bv\bv
F_{Mz}(\bv).
\end{eqnarray}
The third term on the right-hand side of \eq{eq:C_zi_lin_integral_aux} is proportional to an odd moment of $h_{i}$. The rest of the terms are proportional to even moments of $h_{i}$ and they have the same typical size. Grouping terms and recalling \eq{eq:C_zi_lin_differential_2}, we finally obtain expression \eq{eq:C_zi_lin_expanded_in_mass_ratio}.

\section{Eigenfunctions of the operator $\cal K$}
\label{sec:eigenfunctions_of_C_zi_diff}

In this appendix we calculate the eigenfunctions of the operator
\begin{equation}
{\cal K}h := \frac{T}{m_z}
\nabla_v\cdot
\left(
F_{Mz}
\nabla_v
\left(\frac{h}{F_{Mz}}\right)
\right).
\end{equation}

We begin by writing ${\cal K}$ using spherical coordinates $\{v,\beta,\phi\}$ in velocity space, defined in \eq{eq:spherical_coordinates} and the paragraph before it. We are interested in the action of ${\cal K}$ on gyrophase-independent functions, $\partial_\phi g \equiv 0$. Using this, the formulae
\begin{equation}
\nabla_v v = \frac{\bv}{v},
\end{equation}
\begin{equation}
\nabla_v \beta = \frac{1}{v\sin\beta}\left(\cos\beta\frac{\bv}{v} - \bun\right),
\end{equation}
\begin{equation}
\nabla_v \phi = -\frac{1}{v^2\sin^2\beta}\bv\times\bun,
\end{equation}
\begin{equation}
{\cal J} = \frac{1}{(\nabla_v v \times \nabla_v\beta)\cdot\nabla_v\phi} = v^2\sin\beta
\end{equation}
and the expression
\begin{equation}
\fl
\nabla_v\cdot{\bf Y} = \frac{1}{{\cal J}}\left[
\partial_v({\cal J}{\bf Y}\cdot\nabla_v v)
+
\partial_\beta({\cal J}{\bf Y}\cdot\nabla_v \beta)
+
\partial_\phi({\cal J}{\bf Y}\cdot\nabla_v \phi)
\right]
\end{equation}
for the divergence of a vector field, we obtain
\begin{eqnarray}
\fl
{\cal K}h
=
\frac{T}{m_z}
\left[
\frac{1}{v^2}\partial_v\left(v^2 F_{Mz}\partial_v\left(\frac{h}{F_{Mz}}\right)\right)
+
\frac{F_{Mz}}{v^2\sin\beta}\partial_\beta\left(\sin\beta\partial_\beta\left(\frac{h}{F_{Mz}}\right)\right)
\right].
\end{eqnarray}
It is useful to rewrite ${\cal K}$ in coordinates $x := m_z v^2 / (2T)$ and $\beta$. That is,
\begin{eqnarray}
\fl
{\cal K}h
=
2x^{-1/2}\partial_x
\left(
x^{3/2}
F_{Mz}\partial_x\left(\frac{h}{F_{Mz}}\right)\right)
+
\frac{F_{Mz}}{2x\sin\beta}\partial_\beta\left(\sin\beta\partial_\beta\left(\frac{h}{F_{Mz}}\right)\right)
.
\end{eqnarray}
We do not change the notation for $h$ or $F_{Mz}$, although in the previous expression we assume that they are written using $x$ instead of $v$. For example,
\begin{equation}\label{eq:Maxwellian_z_coordinate_x}
F_{Mz} = 
n_z(\rcoor)\left(\frac{m_z}{2\pi T(\rcoor)}\right)^{3/2}
\exp\left(-x\right).
\end{equation}

Take $h = G_l(x) P_l (\cos\beta) F_{Mz}$, where $\{P_l \, \vert \, l=0,1,2\dots\}$ are the Legendre polynomials. These polynomials satisfy the differential equations
\begin{eqnarray}\label{eq:diff_eq_Legendre}
\frac{1}{\sin\beta}\partial_\beta
\left(
\sin\beta\partial_\beta P_l(\cos\beta)
\right)
= -l(l+1) P_l(\cos\beta)
\end{eqnarray}
and the orthogonality relations
\begin{equation}\label{eq:orthogonality_relations_Legendre}
\int_{-1}^{1} P_l(\xi) P_m(\xi)\dd\xi = \frac{2}{2l+1}\delta_{lm}.
\end{equation}

Using \eq{eq:diff_eq_Legendre}, one gets
\begin{eqnarray}
{\cal K}(G_l P_l F_{Mz}) =
({\tilde{\cal K}} G_l)
P_l F_{Mz},
\end{eqnarray}
with
\begin{equation}
{\tilde{\cal K}} G_l = 2x \frac{\dd^2G_l}{\dd x^2} + (3-2x)\frac{\dd G_l}{\dd x} - \frac{l(l+1)}{2x} G_l.
\end{equation}

Let $L_p^{(\alpha)}(x)$, $p = 0,1,2,\dots$, $\alpha\in\mathbb{R}$, denote the generalized Laguerre polynomials. They satisfy the differential equations
\begin{equation}\label{eq:diff_eq_Laguerre}
x \frac{\dd^2 L_p^{(\alpha)}}{\dd x^2} + (\alpha + 1 - x)\frac{\dd L_p^{(\alpha)}}{\dd x} +p L_p^{(\alpha)} = 0
\end{equation}
and the orthogonality relations
\begin{equation}\label{eq:orthogonality_relations_Laguerre}
\int_0^\infty
x^{l+1/2} L_p^{(l+1/2)} L_q^{(l+1/2)} e^{-x}
\dd x
=
\frac{\Gamma(p+l+3/2)}{p!}\delta_{pq}.
\end{equation}
Try $G_l(x) = U_{p,l}(x)$, where $U_{p,l}(x):= x^{l/2} L_p^{(l+1/2)}(x)$. Then,
\begin{equation}
\fl
{\tilde{\cal K}} U_{p,l} =
2 x^{l/2}\left(
x \frac{\dd^2 L_p^{(l+1/2)}}{\dd x^2} + (l+3/2 - x)\frac{\dd L_p^{(l+1/2)}}{\dd x} - \frac{l}{2} L_p^{(l+1/2)}
\right)
.
\end{equation}
Employing \eq{eq:diff_eq_Laguerre}, we deduce that
\begin{equation}
{\tilde{\cal K}} U_{p,l} =
-(l+2p) U_{p,l}
.
\end{equation}
Therefore,
\begin{equation}\label{eq:eigenfunctions}
{\cal K}(U_{p,l} P_l F_{Mz}) = -(l+2p) U_{p,l} P_l F_{Mz}.
\end{equation}

\section{Definitions of $f_c$ and $f_s$}
\label{sec:f_c_and_f_s}

We give here the expressions for $f_c$ and $f_s$, that enter equation \eq{eq:explicit_expression_A0}. These expressions are taken from \cite{Helander2017b}.

The flux function $f_c$ is defined by
\begin{equation}
f_c = \frac{3\langle B^2 \rangle}{4}\int_0^{B_{\rm max}^{-1}} \frac{\lambda}{\langle\sqrt{1-\lambda B}\rangle}\dd\lambda.
\end{equation}
As for $f_s$,
\begin{equation}
f_s = \frac{3\langle B^2 \rangle}{4}\int_0^{B_{\rm max}^{-1}} \frac{\langle g_4 \rangle\lambda}{\langle\sqrt{1-\lambda B}\rangle}\dd\lambda,
\end{equation}
where
\begin{equation}
g_4(\lambda,l) = \sqrt{1-\lambda B} \int_{l_{\rm max}}^{l}
\left(
\bun\times\nabla r
\right)
\cdot
\nabla \left(\frac{1}{\sqrt{1-\lambda B}}\right)
\dd l'
\end{equation}
is defined for $\lambda < B_{\rm max}^{-1}$. The integral is taken along the magnetic field line, $l$ is the arc length along the line and $l_{\rm max}$ gives the position at which $B$ takes the value $B_{\rm max}$.

\section*{References}


\begin{thebibliography}{10}

\bibitem{Burhenn2009}
Burhenn~R \emph{et al.} 2009
\newblock \emph{Nucl. Fusion} {\bf 49} 065005

\bibitem{Putterich2019}
P\"utterich~T, Fable~E, Dux~R, O'Mullane~M, Neu~R and Siccinio~M 2019
\newblock \emph{Nucl. Fusion} {\bf 59} 056013

\bibitem{Braun2010}
Braun~S and Helander~P 2010
\newblock \emph{Phys. Plasmas} {\bf 17} 072514

\bibitem{Helander2017}
Helander~P, Newton~S~L, Moll\'en~A and Smith~H~M 2017
\newblock \emph{Phys. Rev. Lett.} {\bf 118} 155002

\bibitem{Calvo2018b}
Calvo~I, Parra~F~I, Velasco~J~L, Alonso~J~A and Garc\'{\i}a-Rega\~na~J~M 2018
\newblock \emph{Nucl. Fusion} {\bf 58} 124005

\bibitem{Buller2018}
Buller~S, Smith~H~M, Helander~P, Moll\'en~A, Newton~S~L and Pusztai~I 2018
\newblock \emph{J. Plasma Phys.} {\bf 84} 905840409

\bibitem{García-Regaña2013}
Garc\'ia-Rega\~na~J~M, Kleiber~R, Beidler~C~D, Turkin~Y, Maaßberg~H and Helander~P 2013
\newblock \emph{Plasma Phys. Control. Fusion} {\bf 55} 074008

\bibitem{García-Regaña2017}
Garc\'ia-Rega\~na~J~M, Beidler~C~D, Kleiber~R, Helander~P, Moll\'en~A, Alonso~J~A, Landreman~M, Maaßberg~H, Smith~H~M, Turkin~Y and Velasco~J~L 2017
\newblock \emph{Nucl. Fusion} {\bf 57} 056004

\bibitem{Velasco2018}
Velasco~J~L, Calvo~I, Garc\'{\i}a-Rega\~na~J~M, Parra~F~I, Satake~S, Alonso~J~A and the LHD team 2018
\newblock {\em Plasma Phys. Control. Fusion} {\bf 60} 074004

\bibitem{García-Regaña2018}
Garc\'ia-Rega\~na~J~M, Estrada~T, Calvo~I, Velasco~J~L, Alonso~J~A, Carralero~D, Kleiber~R, Landreman~M, Moll\'en~A, S\'anchez~E, Slaby~C, TJ-II Team and W7-X Team.
\newblock \emph{Plasma Phys. Control. Fusion} {\bf 60} 104002

\bibitem{Mollen2018}
Moll\'en~A, Landreman~M, Smith~H~M, Garc\'ia-Rega\~na~J~M and Nunami~M 2018
\newblock {\em Plasma Phys. Control. Fusion} {\bf 60} 084001

\bibitem{Fujita2019}
Fujita~K, Satake~S, Kanno~R, Nunami~M, Nakata~M and Garc\'ia-Rega\~na~J~M 2019
\newblock {Plasma and Fusion Research} {\bf 14} 3403102

\bibitem{Mollen2015}
Moll\'en~A, Landreman~M, Smith~H~M, Braun~S and Helander~P 2015
\newblock {\em Phys. Plasmas} {\bf 22} 112508

\bibitem{Beidler2011}
Beidler~C~D {\it et al} 2011
\newblock {\em Nucl. Fusion} {\bf 51} 076001

\bibitem{Calvo2017}
Calvo~I, Parra~F~I, Velasco~J~L and Alonso~J~A 2017
\newblock \emph{Plasma Phys. Control. Fusion} {\bf 59} 055014

\bibitem{Calvo2018}
Calvo~I, Velasco~J~L, Parra~F~I, Alonso~J~A and Garc\'ia-Rega\~na~J~M 2018
\newblock \emph{J. Plasma Physics} {\bf 84} 905840407

\bibitem{Calvo13}
Calvo~I, Parra~F~I, Velasco~J~L and Alonso~J~A 2013
\newblock {\em Plasma Phys. Control. Fusion} {\bf 55} 125014

\bibitem{Bernstein1985}
Bernstein~I~B and Catto~P~J 1985
\newblock {\em Phys.Fluids} {\bf 28} 1342

\bibitem{Helander2017b}
Helander~P, Parra~F~I and Newton~S~L 2017
\newblock \emph{J. Plasma Phys.} {\bf 83} 905830206

\bibitem{Helander_book_2002}
Helander~P and Sigmar~D~J 2002 {\em Collisional Transport in
Magnetized Plasmas}
\newblock ({\em Cambridge Monographs on Plasma Physics})
\newblock ed Haines~M~G {\em et al}
\newblock (Cambridge, UK: Cambridge University Press)

\bibitem{Boozer81}
Boozer~A~H 1981
\newblock {\em Phys. Fluids} {\bf 24} 1999

\bibitem{Velasco2019}
Velasco~J~L, Calvo~I, Parra~F~I and Garc\'ia-Rega\~na~J~M 2019, ``KNOSOS: a fast orbit-averaging neoclassical code for arbitrary stellarator geometry'', \texttt{arXiv:1908.11615 [physics.plasm-ph]}.

\end{thebibliography}
\end{document}